\documentclass[journal]{IEEEtran}

\usepackage{scalerel}% for \scaleto

\usepackage{graphicx}
\usepackage{amsfonts}
\usepackage{bm}
\usepackage[rgb]{xcolor}

\usepackage{multirow}
\usepackage{makecell}
\usepackage{booktabs}

\usepackage{hyperref}

\usepackage{subcaption}

\usepackage{algorithm}
\usepackage[noend]{algpseudocode}

\newcommand{\VAE}{\textbf{VAE}}
\newcommand{\CDVAE}{\textbf{CDVAE}}
\newcommand{\CDVAESP}{\textbf{CDVAE} [SP]}
\newcommand{\CDVAEMCC}{\textbf{CDVAE} [MCC]}
\newcommand{\CG}{\textbf{CDVAE-GAN}}
\newcommand{\CGS}{\textbf{CDVAE-GAN\textsubscript{SP}}}
\newcommand{\CGM}{\textbf{CDVAE-GAN\textsubscript{MCC}}}
\newcommand{\CGB}{\textbf{CDVAE-GAN\textsubscript{BOTH}}}
\newcommand{\CL}{\textbf{CDVAE-CLS}}

\newcommand{\CLGS}{\textbf{CDVAE-CLS-GAN\textsubscript{SP}}}
\newcommand{\CLGM}{\textbf{CDVAE-CLS-GAN\textsubscript{MCC}}}
\newcommand{\CLGB}{\textbf{CDVAE-CLS-GAN\textsubscript{BOTH}}}

\usepackage{cite}
\usepackage{amsmath}

\begin{document}

\title{Unsupervised Representation Disentanglement using Cross Domain Features and Adversarial Learning in Variational Autoencoder based Voice Conversion}

\author{Wen-Chin~Huang,~\IEEEmembership{Student Member,~IEEE,} Hao~Luo, Hsin-Te~Hwang, Chen-Chou~Lo, Yu-Huai~Peng, Yu~Tsao,~\IEEEmembership{Member,~IEEE,}
        and~Hsin-Min~Wang,~\IEEEmembership{Senior Member,~IEEE}% <-this % stops a space
        
\thanks{\copyright 2020 IEEE. Personal use of this material is permitted. Permission from IEEE must be obtained for all other uses, in any current or future media, including reprinting/republishing this material for advertising or promotional purposes, creating new collective works, for resale or redistribution to servers or lists, or reuse of any copyrighted component of this work in other works.}% <-this % stops a space
\thanks{Wen-Chin Huang is with the Graduate School of Informatics, Nagoya University, Japan. This work was done while he was with the Institute of Information Science, Academia Sinica, Taipei, Taiwan. e-mail: wen.chinhuang@g.sp.m.is.nagoya-u.ac.jp.}% <-this % stops a space
\thanks{Hao~Luo, Hsin-Te~Hwang, Chen-Chou~Lo, Yu-Huai~Peng and Hsin-Min~Wang are with the Institute of Information Science, Academia Sinica, Taipei, Taiwan.}% <-this % stops a space
\thanks{Yu Tsao is with the Research Center of Information Technology Institute of Information Science, Academia Sinica, Taipei, Taiwan.}% <-this % stops a space
}

\maketitle

\begin{abstract}
An effective approach for voice conversion (VC) is to disentangle linguistic content from other components in the speech signal. The effectiveness of variational autoencoder (VAE) based VC (VAE-VC), for instance, strongly relies on this principle. In our prior work, we proposed a cross-domain VAE-VC (CDVAE-VC) framework, which utilized acoustic features of different properties, to improve the performance of VAE-VC. We believed that the success came from more disentangled latent representations. In this paper, we extend the CDVAE-VC framework by incorporating the concept of adversarial learning, in order to further increase the degree of disentanglement, thereby improving the quality and similarity of converted speech. More specifically, we first investigate the effectiveness of incorporating the generative adversarial networks (GANs) with CDVAE-VC. Then, we consider the concept of domain adversarial training and add an explicit constraint to the latent representation, realized by a speaker classifier, to explicitly eliminate the speaker information that resides in the latent code. Experimental results confirm that the degree of disentanglement of the learned latent representation can be enhanced by both GANs and the speaker classifier. Meanwhile, subjective evaluation results in terms of quality and similarity scores demonstrate the effectiveness of our proposed methods.
\end{abstract}

\begin{IEEEkeywords}
voice conversion, unsupervised learning, disentangled representation, variational autoencoder, adversarial learning, cross domain features.
\end{IEEEkeywords}

\section{Introduction}

\IEEEPARstart{V}{oice} conversion (VC) aims to convert the speech from a source to that of a target without changing the linguistic content \cite{vc-survey}. 
Speaker voice conversion \cite{first-speaker-conversion} is a typical type of VC and refers to the process of converting speech from a source speaker to a target speaker. In addition, a wide variety of applications could be solved by applying VC, such as accent conversion \cite{accent-conversion}, personalized speech synthesis \cite{personalized-tts, personalized-expresive-tts}, and speaking-aid device support \cite{EL-GMM, EL-o2m, EL-hybrid}. Since the spectral property plays an important role in characterizing speaker individuality, spectral conversion has been intensively studied in VC. In this work, we focus on spectral mapping in speaker voice conversion.

Numerous VC approaches have been proposed. The Gaussian mixture model (GMM)-based method \cite{VC,GMM-VC} has been a popular statistical approach that estimates the joint density of the source-target feature vectors, which requires a training procedure and has a well-known disadvantage that the converted outputs generally suffer from an over-smoothing issue. Frequency warping methods, such as vocal tract length normalization \cite{VTLN-VC}, weighted frequency warping \cite{weighted-freq-warping} and dynamic frequency warping \cite{dynamic-freq-warping}, are able to keep spectral details while providing inferior speaker identity conversion quality to that of statistical approaches. Exemplar-based methods \cite{exemplar-noisy-VC,exemplar-residual-VC,LLE-VC, sparse-rep, exemplar-FW-VC} require much less training data and are capable of modeling the high-dimensional spectra. In recent years, deep neural networks (DNNs) have established supremacy in a wide range of research fields, including VC \cite{ANN-VC-first, ANN-VC, layerwise-VC, VC-DBLSTM}. DNNs have been utilized for not only spectral mapping but also neural vocoding \cite{wavenet, SDWN, SIWN}. It has been shown that employing neural vocoders as the waveform generation module can greatly improve the performance of VC systems \cite{VCWN, SAWN-Sisman, SAWN-iFLYTEK, SAWN-PATRICK, VAE-WNV-FT, SAWN-GAN}. It has also been shown that VC systems, whether implemented in high-dimensional or low-dimensional features, benefit from spectral detail compensation \cite{exemplar-residual-VC, exemplar-FW-VC, VC-LLE-compensation}.

Nonetheless, most of the approaches described above rely on the availability of parallel training data, which is often not accessible in real world scenarios. Thus, the development of non-parallel VC methods has been gaining attention \cite{vcc2018}. One approach is to construct a pseudo parallel dataset from a non-parallel corpus \cite{INCA}. Another family of approaches utilizes a pre-trained automatic speech recognition model to compute the phonetic posteriorgram (PPG) as the speaker-independent linguistic feature, followed by a PPG-to-acoustic mapping to generate converted features \cite{VC-PPG, VAE-PPG-DVEC-VC}. A recently popular approach is to use DNNs to model the probability distribution of the target features; state-of-the-art models such as variational autoencoders (VAEs) \cite{VAE} and generative adversarial networks (GANs) \cite{GAN} have been successfully applied to non-parallel VC  \cite{VAE-VC, VAE-PPG-DVEC-VC, ACVAE-VC, VAE-GAN-VC, cyclegan-vc, cyclegan-vc-2, stargan-vc, CHOU-NPVC, can-we-steal}.

In this work, we focus on VAE-based VC (VAE-VC) \cite{VAE-VC}. Specifically, the spectral conversion function is composed of an encoder-decoder pair. The encoder encodes the input spectral feature into a latent code; the decoder mixes the latent code and a specified target speaker code to generate the converted feature. The encoder-decoder network and the speaker codes are trained by back-propagation of the reconstruction error, along with a Kullback-Leibler (KL)-divergence loss that regularizes the distribution of the latent code.

\begin{figure}[t]
  \centering
  \includegraphics[width=0.48\textwidth]{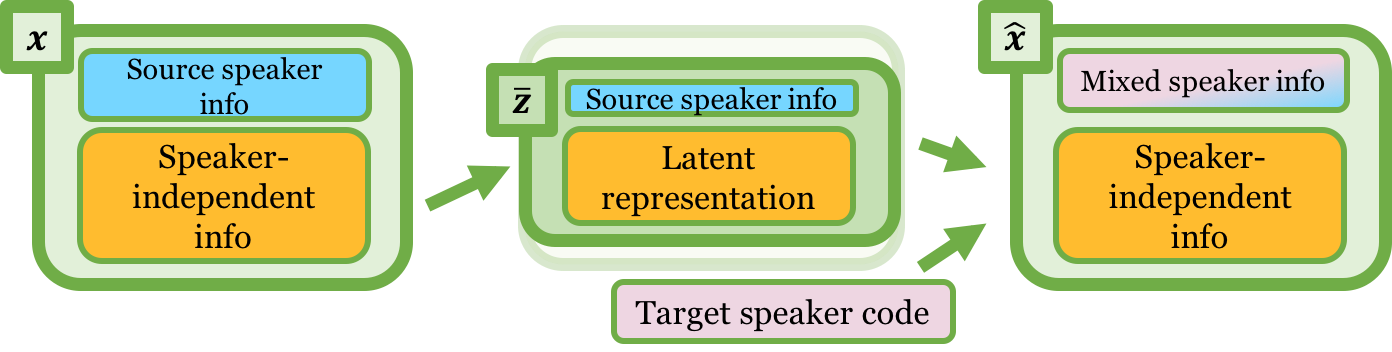}
  \centering
  \caption{Illustration of how entangled latent representation affects the conversion performance in a general VAE-VC framework. The residual source speaker information in the latent code will be mixed with the given target speaker code, resulting in a mixed speaker identity in the converted feature. Thus, the performance might be harmed.
  }
  \label{fig:entangled-vc}
\end{figure}

\begin{figure*}[t]
  \centering
  \includegraphics[width=0.75\textwidth]{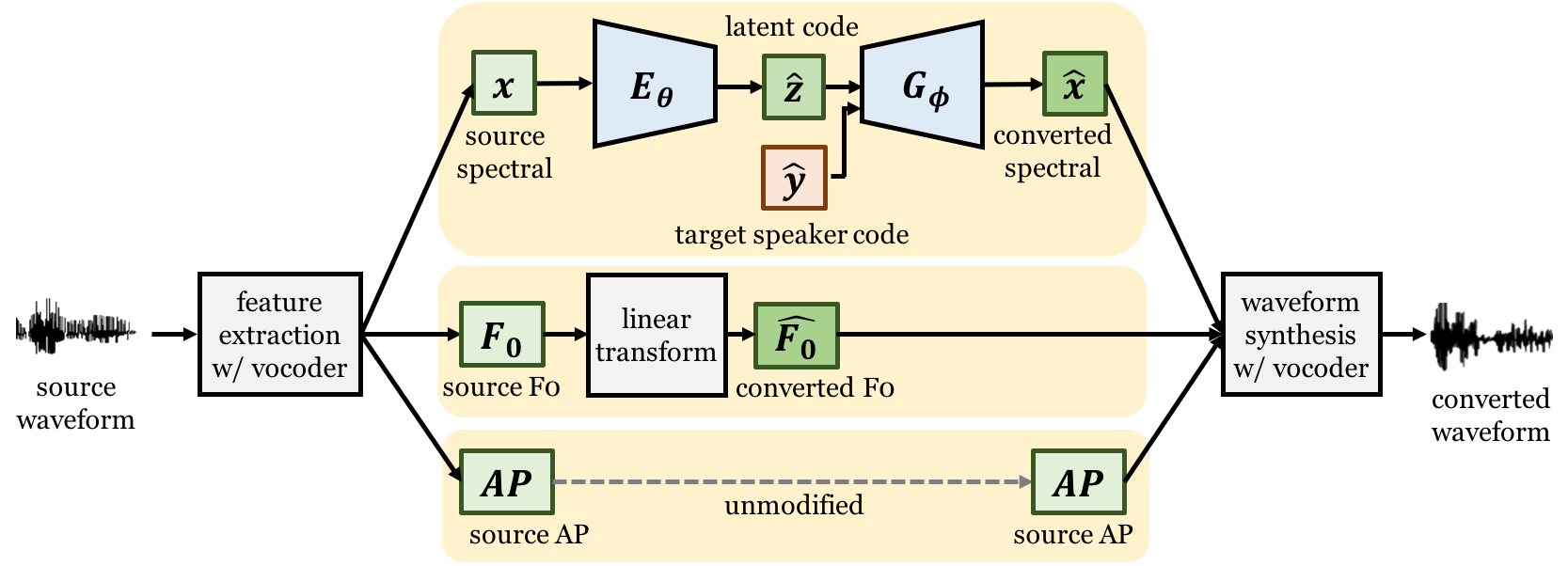}
  \centering
  \caption{Illustration of the conversion phase of the VAE-VC \cite{VAE-VC} framework. Following traditional VC systems, a vocoder first parameterizes the waveform into acoustic features, which are then converted in different streams, and finally the converted features are used to synthesize the converted waveform by a vocoder.}
  \label{fig:vae-vc}
\end{figure*}

The degree of disentanglement of the latent representation is crucial to the success of many speech processing frameworks \cite{FHVAE, GMVAE, GMVAE-adversarial, speech2vec, Representation-WNV}, including VAE-VC. Since we focus on the task of speaker voice conversion, the degree of disentanglement is defined as the amount of (source) speaker information residing in the latent code, i.e., the independence of the latent code and the speaker code \cite{disentanglement}. An illustration is given in Figure~\ref{fig:entangled-vc}. If the latent code is entangled by multiple components (e.g., in the VC task, the source speaker information remains in the latent code), during conversion, the decoder will draw the speaker information from both the given target speaker code and the residual source speaker information in the latent code, which harms the conversion performance. From the success of VAE-VC, we can infer that, at least to some extent, the decoder is trained to use more information in the given speaker code, rather than the speaker characteristics remained in the latent code, otherwise conversion made by changing the speaker code will not work. Although the success may be a natural result of model optimization, we doubt whether the performance is robust enough. For instance, in \cite{autovc}, it was demonstrated that the performance of autoencoder-based VC models was sensitive to the latent space dimension. This raises the need to design better schemes for making the latent code more independent of the speaker.

In our prior work \cite{CDVAE}, we proposed a cross-domain VAE-based VC framework (referred to as CDVAE-VC in the following discussion). The motivations of CDVAE-VC are: (1) although the effectiveness of VAE-VC using vocoder spectra (e.g., the STRAIGHT spectra, SPs \cite{STRAIGHT}) has been confirmed, the use of other types of spectral features, such as mel-cepstral coefficients (MCCs) \cite{MCC} that are related to human perception and have been widely used in VC, have not been properly investigated; (2) since modeling the low- and high-dimensional features alone has their respective shortcomings, based on multi-target/task learning \cite{Multitarget, Multitask}, it is believed that a model capable of simultaneously modeling two types of spectral features can yield better performance even if they are from the same feature domain. To this end, CDVAE-VC \cite{CDVAE} extended the VAE-VC framework to jointly consider two kinds of spectral features, namely SPs and MCCs. By introducing two additional cross-domain reconstruction losses and a latent similarity constraint into the training objective, the latent representations encoded from the input SPs and MCCs are biased to each other and capable of self- or cross-reconstructing the input features. We speculated that the success of CDVAE-VC came from the fact that a more disentangled latent representation was learned. Furthermore, we observed a positive correlation between the conversion performance and the extent to which the latent code was disentangled.

In this work, we extend the CDVAE-VC framework by incorporating the concept of adversarial training to improve the degree of disentanglement as well as the conversion performance. First, we directly combine CDVAE-VC with GANs. GANs have shown the ability to enhance the output of the decoder in encoder-decoder network based VC frameworks \cite{CHOU-NPVC}. Therefore, it is expected that such a combination can improve the quality of converted speech. Second, inspired from the idea of domain adversarial training (DAT) \cite{DAT}, we add a speaker classification training objective to the latent variables, in order to explicitly project away speaker-related information. A similar idea has been applied to several speech processing tasks, such as speech recognition \cite{DAT-ASR, DAT-ASR-adaptation, DAT-ASR-accent}, speech enhancement \cite{DAT}, VC \cite{CHOU-NPVC, AE-NPVC} and singing VC \cite{USVC}. Here, we utilize DAT by considering cross-domain features to further facilitate a more disentangled latent representation.

Designing a clear evaluation metric for degree of disentanglement has long been an open problem in the field of machine learning. In image modeling, visual inspection has been a standard and intuitive approach \cite{tcvae, DOD-quant}. However, the visual inspection is not perfectly feasible for speech processing tasks since it is hard to quantify the difference in voices as a specific latent variable changes. In previous works \cite{CHOU-NPVC, CHOU-M2MVC, autovc}, a classifier-based metric has been proposed. Since the metric is also based on a trained classifier, it has limitations in comparing the disentanglement between different latent codes obtained by different models due to different training conditions and dynamics. Following \cite{F0-FCN-CDVAE}, we utilize the parallel data that exist in most benchmark VC datasets and derive a novel metric for measuring disentanglement. The key assumption is that an ideal encoder should encode a pair of parallel sentences uttered by two different speakers to similar latent codes. We measure the cosine similarity between such latent codes to evaluate how well the encoder disentangles the latent codes.

The remainder of this paper is organized as follows. In Section~\ref{sec:background}, we first review the VAE-VC and its extended version, CDVAE-VC. Section~\ref{sec:GAN} introduces how to combine GANs with CDVAE-VC. Then, we describe how to add an adversarial speaker classifier objective to the latent code in Section~\ref{sec:cls}. In Section~\ref{sec:experiments}, we first examine our proposed mechanisms one by one, using conventional objective and subjective evaluation metrics adopted in VC. Disentanglement measurements of our proposed methods and how they are related to the VC performance are presented afterwards. Finally, we conclude the paper with discussions in Section~\ref{sec:conclusion}.

\section{Background}
\label{sec:background}

In conventional VC frameworks, the acoustic features of the source speaker are converted to those of the target speaker in different feature streams. Many researches focus on the conversion of spectral features \cite{GMM-VC} and thus formulate VC as follows. Given $N$ source speaker's spectral frames $\mathbf{X}_s=\{\bm{x}_{s,1},\dots,\bm{x}_{s,N}\}$, the goal is to find a conversion function $f$ such that
\begin{equation}
\label{eq:formulation}
	\hat{\bm{x}}_{t,n}=f(\bm{x}_{s,n}).
\end{equation}
Note that the second subindices in both sides of the equation are both $n$, which means that the converted spectral feature sequence has the same length with that of the source. In the rest of the article, we drop the frame or the speaker indices for simplicity.

In the following subsections, we describe two VAE based VC frameworks. Throughout the paper, we use ``bar" to indicate the reconstructed features, and ``hat" to indicate the converted features.

\subsection{VAE-VC}
\label{ssec:vae-vc}

Figure~\ref{fig:vae-vc} depicts the conversion process of a typical VAE-VC system \cite{VAE-VC}. The core of VAE-VC is an encoder-decoder network. During training, given an observed (source or target) spectral frame $\bm{x}$, a speaker-independent encoder $E_\theta$ with parameter set $\theta$ encodes $\bm{x}$ into a latent code: $\bar{\bm{z}}=E_\theta(\bm{x})$. The speaker code $\bm{y}$ of the input frame is then concatenated with the latent code, and passed to a conditional decoder $G_\phi$ with parameter set $\phi$ to reconstruct the input. This reconstruction process can be expressed as:
\begin{equation}
	\label{eq:self-recon}
	\bar{\bm{x}}=G_\phi(\bar{\bm{z}},\bm{y})=G_\phi(E_\theta(\bm{x}),\bm{y}).
\end{equation}

The model parameters can be obtained by maximizing the variational lower bound:
\begin{align}
	\mathcal{L}_{vae}(\theta,\phi;\bm{x}, \bm{y}) &= \mathcal{L}_{recon}(\bm{x},\bm{y})+\mathcal{L}_{lat}(\bm{x}), \label{eq:vae_loss} \\
	\mathcal{L}_{recon}(\bm{x},\bm{y}) &= \mathbb{E}_{\bm{z}\sim q_\theta(\bar{\bm{z}}|\bm{x})}\bigl[\log p_\phi(\bar{\bm{x}}|\bm{z},\bm{y})\bigr],\\
	\mathcal{L}_{lat}(\bm{x}) &= -D_{KL}(q_\theta(\bar{\bm{z}}|\bm{x}) \Vert p(\bm{z})),
\end{align}
where $q_\theta(\bar{\bm{z}}|\bm{x})$ is the approximate posterior, $p_\phi(\bar{\bm{x}}|\bm{z},\bm{y})$ is the data likelihood, and $p(\bm{z})$ is the prior distribution of the latent space. $\mathcal{L}_{recon}$ is simply a reconstruction term as in any vanilla autoencoder, whereas $\mathcal{L}_{lat}$ regularizes the encoder to align the approximate posterior with the prior distribution.

In the conversion phase, one could use \eqref{eq:self-recon} to formulate the conversion function $f$:
\begin{equation}
	\label{eq:conversion}
	\hat{\bm{x}}=f(\bm{x}, \hat{\bm{y}})=G_\phi(\hat{\bm{z}},\hat{\bm{y}})=G_\phi(E_\theta(\bm{x}),\hat{\bm{y}}),
\end{equation}
where $\hat{\bm{y}}$ is the target speaker code.

The VAE framework makes several assumptions. First, $p_\phi(\bar{\bm{x}}|\bm{z},\bm{y})$ is assumed to follow a normal distribution whose covariance is an identity matrix. Second, $p(\bm{z})$ is set to be a standard normal distribution. Third, the expectation over $\bm{z}$ is approximated by sampling via a linear-transformation based re-parameterization trick \cite{VAE}. With these simplifications, we can avoid intractability and optimize the autoencoder parameter sets $\theta\cup\phi$ and the speaker codes via back-propagation.

\begin{figure}[t]
  \centering
  \includegraphics[width=0.35\textwidth]{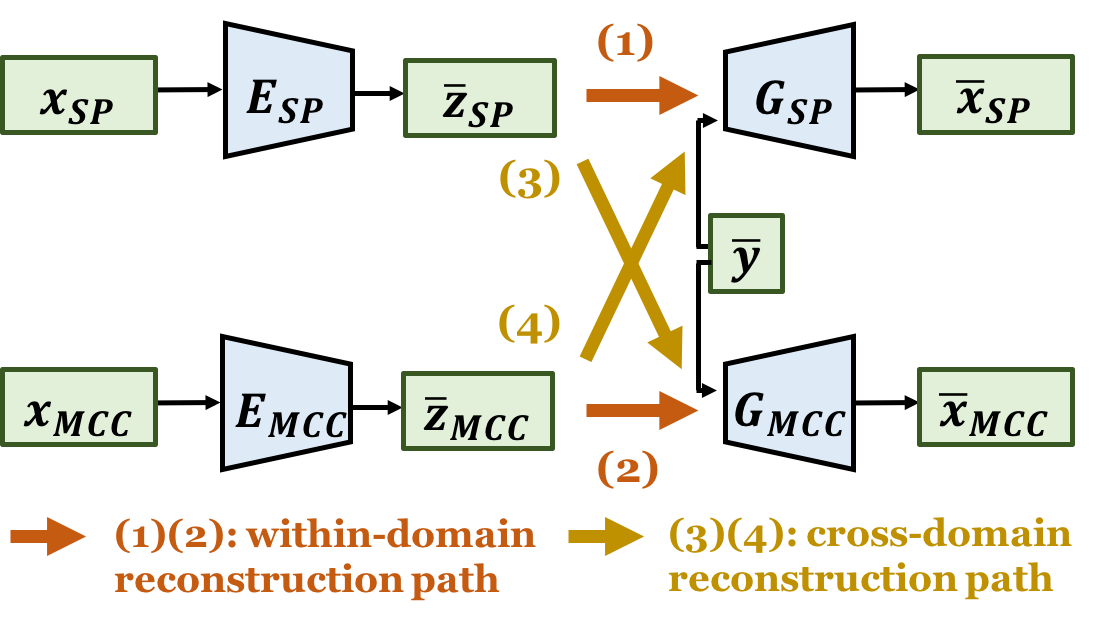}
  \centering
  \caption{Illustration of the training phase of the CDVAE-VC \cite{CDVAE} framework. In this framework, each feature has its own set of encoder and decoder. During training, by minimizing the loss derived from the within- and cross-domain reconstruction paths, the latent codes $\bm{z_{SP}}$ and $\bm{z_{MCC}}$ learn to reconstruct not only corresponding input features but also the cross-domain features.}
  \label{fig:cdvae-vc}
\end{figure}

\subsection{CDVAE-VC}
\label{ssec:cdvae}

In \cite{CDVAE}, we proposed the CDVAE-VC framework to utilize spectral features of different properties extracted from the same observed speech frame. As depicted in Figure~\ref{fig:cdvae-vc}, the CDVAE framework is formed by a collection of encoder-decoder pairs, one for each kind of spectral feature. Considering the SPs and MCCs as two kinds of spectral features (denoted as $\bm{x}_{\scaleto{SP}{3pt}}$ and $\bm{x}_{\scaleto{MCC}{3pt}}$ ), the following losses are defined:
\begin{align}
	\bar{\bm{z}}_{\scaleto{SP}{3pt}}&=E_{\scaleto{SP}{3pt}}(\bm{x}_{\scaleto{SP}{3pt}}), \bar{\bm{z}}_{\scaleto{MCC}{3pt}}=E_{\scaleto{MCC}{3pt}}(\bm{x}_{\scaleto{MCC}{3pt}}),\\
	\bar{\bm{x}}_{\scaleto{S-S}{3pt}}&=G_{\scaleto{SP}{3pt}}(\bar{\bm{z}}_{\scaleto{SP}{3pt}},\bm{y}), \bar{\bm{x}}_{\scaleto{M-M}{3pt}}=G_{\scaleto{MCC}{3pt}}(\bar{\bm{z}}_{\scaleto{MCC}{3pt}},\bm{y}),\\
	\bar{\bm{x}}_{\scaleto{S-M}{3pt}}&=G_{\scaleto{MCC}{3pt}}(\bar{\bm{z}}_{\scaleto{SP}{3pt}},\bm{y}), \bar{\bm{x}}_{\scaleto{M-S}{3pt}}=G_{\scaleto{SP}{3pt}}(\bar{\bm{z}}_{\scaleto{MCC}{3pt}},\bm{y}),\\
	\mathcal{L}_{in}&=\mathcal{L}_{recon}(\bar{\bm{x}}_{\scaleto{S-S}{3pt}},\bm{y})+\mathcal{L}_{recon}(\bar{\bm{x}}_{\scaleto{M-M}{3pt}},\bm{y}),\\
	\mathcal{L}_{KLD}&=\mathcal{L}_{lat}(\bm{x}_{\scaleto{SP}{3pt}})+\mathcal{L}_{lat}(\bm{x}_{\scaleto{MCC}{3pt}}),\\
	\mathcal{L}_{cross}&=\mathcal{L}_{recon}(\bar{\bm{x}}_{\scaleto{S-M}{3pt}},\bm{y})+\mathcal{L}_{recon}(\bar{\bm{x}}_{\scaleto{M-S}{3pt}},\bm{y}),
\end{align}
where $E_{\scaleto{SP}{3pt}}$ and $G_{\scaleto{SP}{3pt}}$ are the encoder and decoder for SPs, and $E_{\scaleto{MCC}{3pt}}$ and $G_{\scaleto{MCC}{3pt}}$ are the encoder and decoder for MCCs; $\bar{\bm{x}}_{\scaleto{S-S}{3pt}}$ and $\bar{\bm{x}}_{\scaleto{M-M}{3pt}}$, respectively, denote the generated SPs and MCCs from the within-domain reconstruction paths; $\bar{\bm{x}}_{\scaleto{M-S}{3pt}}$ and $\bar{\bm{x}}_{\scaleto{S-M}{3pt}}$, respectively, denote the generated SPs and MCCs from the cross-domain reconstruction paths. Note that $\mathcal{L}_{recon}(\cdot,\bm{y})$ calculates the reconstruction loss between the first argument and the corresponding input feature.

In short, we introduce two extra reconstruction streams.  By minimizing the cross-domain reconstruction loss, we enforce $\bm{z}_{\scaleto{SP}{3pt}}$ to contain enough information to reconstruct $\bm{x}_{\scaleto{MCC}{3pt}}$, and vice versa. As a result, the behavior of the encoders for both feature domains are constrained to be the same, i.e., they are expected to extract similar latent information from different types of input spectral features. To explicitly reinforce this constraint, a latent similarity L1 loss defined as
\begin{align}
	\mathcal{L}_{sim}&=-\Vert \bar{\bm{z}}_{\scaleto{SP}{3pt}}-\bar{\bm{z}}_{\scaleto{MCC}{3pt}} \Vert _1 ,
\end{align}
can be included in the final objective expressed as:
\begin{equation}
	\label{eq:cdvae_objective}
	\mathcal{L}_{cdvae}=\mathcal{L}_{in}+\mathcal{L}_{KLD}+\mathcal{L}_{cross}+\mathcal{L}_{sim}.
\end{equation}

The model parameters can be learned by maximizing \eqref{eq:cdvae_objective}. In the conversion phase, there are four conversion paths (i.e., two within-domain and two cross-domain paths). As reported in \cite{CDVAE}, the CDVAE MCC-MCC path gave the best performance in terms of subjective evaluation, which matched the assumption that MCCs are more related to human perception.

\begin{figure*}[t]
  \centering
  \includegraphics[width=\textwidth]{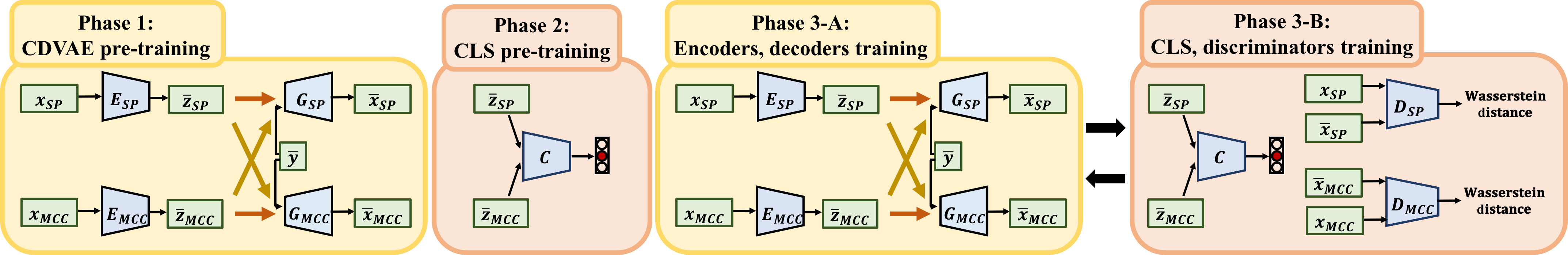}
  \centering
  \caption{Illustration of the training procedure of our proposed CDVAE-CLS-GAN model. \textit{Phase 1:} A CDVAE is trained. \textit{Phase 2:} The latent codes are used to train the CLS. \textit{Phase 3-A and 3-B:} The encoders, decoders and the CLS, discriminators are trained in an alternating order.}
  \label{fig:proposed}
\end{figure*}

\section{Incorporating CDVAE-VC with GANs}
\label{sec:GAN}

Minimizing the reconstruction loss in VAE-VC and CDVAE-VC tends to result in blurry spectra, similar to the over-smoothing effects in other VC frameworks. It is expected that introducing a GAN objective \cite{GAN} can guide the output spectra to be more realistic. In this section, we present the main concepts and system architectures of the combination of GANs and the VAE-VC and CDVAE-VC frameworks.

\subsection{The GAN objective in the general VAE-VC}
\label{ssec:VAE-GAN}

We follow \cite{VAEGAN} and incorporate a GAN objective into the decoder in the original VAE-VC. Assume that the real data distribution of any spectral frame admits density $p^*$, and the autoencoding process defined in \eqref{eq:self-recon} induces a conditional probability $p_{\bar{\bm{x}}}$. From the data distribution prospective of view, the goal is to enhance the decoder network $G$ in \eqref{eq:self-recon} such that $p_{\bar{\bm{x}}}$ best approximates the real data distribution $p^*$:
	\begin{equation}
		p_{\bar{\bm{x}}}(G_\phi(E_\theta(\bm{x}),\bm{y})) \approx p^*(\bm{x}).
	\end{equation}
		
A typical GAN \cite{GAN} realizes the above-mentioned probability approximation by introducing a discriminator $D_{\psi}$ with parameter set $\psi$ that judges whether an input follows a true and natural probability distribution or an artificial one. Together with a generator $G$ that tries to produce realistic output features, these two components play a min-max game and seek an equilibrium with the Jensen-Shannon divergence $\mathcal{D}_{JS}$ as the objective, which is defined as follows:
	\begin{multline}
		\mathcal{L}_{gan}(\theta, \psi; x)=2\,\mathcal{D}_{JS}(p^* \Vert p_{\bar{\bm{x}}})+2\log 2 \\
		=\mathbb{E}_{\bm{x}\sim p^*}\bigl[\log D_\psi(\bm{x})\bigr]+\mathbb{E}_{\bm{z}\sim q_\theta}\bigl[\log D_\psi(G_\phi(\bm{z}) \bigr].
	\end{multline}

To facilitate stable training, in this work we adopt a Wasserstein GAN (WGAN) \cite{WGAN, WGAN-GP}. In the WGAN, the following Wasserstein distance is derived:
	\begin{equation}
		W(p_*,p_{\bar{\bm{x}}})=\sup_{\|D\|_L\leq 1}
		\mathbb{E}_{\bm{x}\sim p^*}\lbrack D(\bm{x})\rbrack -\mathbb{E}_{\bm{x}\sim p_{\bar{\bm{x}}}}\lbrack D(\bm{x})\rbrack,
	\end{equation}
where the supremum is over all 1-Lipschitz functions $D:\mathcal{X}\to \mathbb{R}$. Based on the above distance, the following WGAN loss can be defined:
	\begin{equation}
		\mathcal{L}_{wgan}(\bm{x})=\mathbb{E}_{\bm{x}\sim p^*}\lbrack D_\psi(\bm{x})\rbrack -
		\mathbb{E}_{\bm{z}\sim q_\theta(\bar{\bm{z}}|\bm{x})}\lbrack D_\psi(G_\phi(\bm{z}, \bm{y}))\rbrack,
		\label{eq:wgan_loss}
	\end{equation}
where $D_\psi$ is now a 1-Lipschitz discriminator. Finally, we can combine the objectives of VAE and WGAN by assigning the decoder of VAE as the generator of WGAN. As a result, combining the WGAN loss \eqref{eq:wgan_loss} and the VAE loss \eqref{eq:vae_loss} results in a VAEGAN objective:
	\begin{multline}
		\label{eq:vaegan_loss}
		\mathcal{L}_{vaegan}(\theta,\phi,\psi;\bm{x}, \bm{y}) = \mathcal{L}_{vae}(\bm{x},\bm{y})+ \alpha\mathcal{L}_{wgan}(\bm{x}),	
	\end{multline}
	where $\alpha$ is the weight of the WGAN loss. This objective is shared across the encoder, decoder, and  discriminator. As in standard GAN training, the discriminator is first updated by maximizing this objective, and the encoder and decoder are updated by minimizing the objective. Therefore, the components are optimized in an alternating order. GANs produce more realistic (in our case, sharper) outputs because they optimize a loss function between two distributions in a more direct fashion.

The VAW-GAN-VC method in \cite{VAE-GAN-VC} has a similar motivation to better model spectral features to improve feature generation. However, there is a fundamental difference between the training procedures of VAW-GAN-VC and the training procedures here. In VAW-GAN-VC, the objective of WGANs is to minimize the Wasserstein distance of the two distributions of the converted features and the real target features. Although this is a strong objective, it also brings some limitations. The original VAE-VC and CDVAE-VC consider only auto-encoding in the training phase, and perform conversion by changing the speaker code in the conversion phase. In other words, multiple conversion pairs are integrated into one model, sometimes referred to as ``multi-target" training in VC. VAW-GAN-VC, in contrast, needs to consider not only auto-encoding but also conversion in the training phase, since the discriminator needs to discriminate the real target features and the converted features in order to align the distribution of the latter to that of the former. As a result, VAW-GAN-VC is trained to convert from one source to one target, which limits the flexibility of the model. In this work, we intend to maintain the multi-target flexibility in CDVAE-VC and thus design the WGAN objective to match the distributions of the real features and the reconstructed features. Considering this fundamental difference and to avoid confusion, we focus on multi-target VC and do not take VAW-GAN-VC into discussion and comparison in this paper.

\subsection{CDVAE-VC with GANs (CDVAE-GAN)}
\label{ssec:CDVAE-GAN}

Now we can combine the GAN objective with CDVAE-VC, which we will refer to as CDVAE-GAN, where the derivation of the objective is as simple as replacing the VAE loss in \eqref{eq:vaegan_loss} with the CDVAE objective defined in \eqref{eq:cdvae_objective}. However, in practice, combining CDVAE-VC with GANs is not as trivial as replacing the encoder and decoder in VAE-GAN with CDVAE. For each kind of feature, a separate discriminator should be trained, i.e., $D_{\scaleto{SP}{3pt}}$ and $D_{\scaleto{MCC}{3pt}}$ should be considered. It seems natural to train two discriminators jointly with the whole network. However, as mentioned above, the MCC-MCC path in CDVAE-VC performs best in four paths in the conversion phase. Introducing a discriminator for SPs might not necessarily benefit the quality of the output MCCs. To determine the best architecture, we examine the effect of three settings, including combining CDVAE with only $D_{\scaleto{SP}{3pt}}$, only $D_{\scaleto{MCC}{3pt}}$, and both $D_{\scaleto{SP}{3pt}}$ and $D_{\scaleto{MCC}{3pt}}$. Detailed experimental results will be shown in Sections~\ref{ssec:exp-gan-features} and \ref{ssec:exp-gan}.

\section{Adversarial speaker classifier (CLS)}
\label{sec:cls}

As discussed above, the viability of the family of VAE-VC frameworks relies on the decomposition of input, which is assumed to be composed of phonetic representation and speaker information. Ideally, the latent code extracted using the encoder should contain solely phonetic information and free from any speaker information. However, this decomposition is not explicitly guaranteed. To this end, we investigate the effect of an adversarial speaker classifier to explicitly force the latent code to be speaker independent.

\subsection{The classifier loss}
\label{ssec:cls-loss}

An adversarial speaker classifier $C_\Psi$ with parameter set $\Psi$ tries to classify which speaker the latent code comes from. We will refer to this classifier as CLS. Specifically, given a latent code $\bm{z}$, the CLS predicts a posterior probability $P(\bm{y}=y|\bm{z})$, which is the probability that $\bm{z}$ is extracted from an input frame produced by speaker $y$. Therefore, we can define the CLS loss as the negative cross-entropy between the predicted posterior and the one-hot ground truth vector:
\begin{equation}
	\label{eq:cls_loss}
	\mathcal{L}_{cls}(\bm{x}, \bm{y})=\mathbb{E}_{\bm{z}\sim q_\theta(\bar{\bm{z}}|\bm{x})}\bigl[-\log P(\bm{y}|\bm{z})\bigr].
\end{equation}

\subsection{CDVAE-GAN with CLS (CDVAE-CLS-GAN)}
\label{ssec:CDVAE-CLS-GAN}

We now augment the CDVAE-GAN framework with the adversarial speaker classifier, which we will refer to as CDVAE-CLS-GAN. Adding the CLS loss \eqref{eq:cls_loss} to the CDVAE-GAN loss, we obtain the final objective:
	\begin{multline}
		\label{cdvae-cls-gan-loss}
		\mathcal{L}_{all}(\theta,\phi,\psi,\Psi;\bm{x}, \bm{y}) = \mathcal{L}_{cdvae}(\bm{x},\bm{y})\\+ \alpha\mathcal{L}_{wgan}(\bm{x}) + \lambda \mathcal{L}_{cls}(\bm{x}, \bm{y}),	
	\end{multline}
	where $\lambda$ is the weight of the classifier loss. This objective is shared, again, across the encoder, decoder, discriminator, and classifier.

 The training process is divided into three phases, as depicted in Figure~\ref{fig:proposed}. Phase one involves the training of the VAE. In phase two, to pre-train the classifier, we first use the trained VAE obtained in phase one to extract latent codes from the same training set. The classifier is then trained with these latent codes to minimize \eqref{eq:cls_loss}. In the third phase, we train the whole network using an alternating update schedule, similar to the one described in Section~\ref{ssec:VAE-GAN}. Specifically, the encoder and the decoder are first frozen and the discriminator and classifier are trained to \textit{maximize} $\mathcal{L}_{wgan}$ and $\textit{minimize}$ $\mathcal{L}_{cls}$ defined in \eqref{eq:wgan_loss} and \eqref{eq:cls_loss}, respectively, and thus they can discriminate self-reconstructed features and classify latent codes correctly. Then, we freeze these modules and train the encoder and decoder to not only minimize $\mathcal{L}_{cdvae}$ in \eqref{eq:cdvae_objective}, but also optimize $\mathcal{L}_{wgan}$ and $\mathcal{L}_{cls}$ so that they can fool the frozen components. 

The described training scheme also plays a min-max game between \{encoders, decoders\} and \{discriminator, classifier\}. An ideally trained model should contain encoders that learns to project away as much speaker information as possible and decoders that can generate realistic and natural output spectra given an inferred latent code with a specific speaker code. Algorithm~\ref{algo1} summarizes the training procedure of CDVAE-CLS-GAN. 

    \begin{algorithm}[t]
    \caption{Training procedure of CDVAE-CLS-GAN}
      \begin{algorithmic}[0]
      \Function{autoencode}{$\bm{x}, \bm{y}$}
        \State $\bm{z} \leftarrow$ sample using $E(\bm{x})$
        \State $\bar{\bm{x}} \leftarrow G(\bm{z},\bm{y})$
        \State \Return $\bar{\bm{x}},~\bm{z}$
      \EndFunction  
      \State
      
      \State $\phi, \theta, \psi, \Psi ~\leftarrow$ initialization
      
      \State // Phase 1: train the VAE
      \While {not converged}
        \State $X, Y \leftarrow$ mini-batch of samples from the training set
        \State $\bar{X}, Z \leftarrow$ AUTOENCODE($X, Y$)
        \State $\phi, \theta ~ \xleftarrow{update} ~ -\nabla \mathcal{L}_{cdvae}(\bar{X}, Y)$
      \EndWhile
      \State
      
      \State // Phase 2: train the CLS
      \While {not converged}
        \State $X, Y \leftarrow$ mini-batch of samples from the training set
        %\State $\bar{Z}$ $\leftarrow$ E($X$)
        \State $\mathcal{L}_{cls} \leftarrow \mathcal{L}_{cls}(X,Y)$
        \State $\Psi ~ \xleftarrow{update} ~ -\nabla \mathcal{L}_{cls}(X, Y)$
      \EndWhile
      \State
      
      \State // Phase 3: train the whole network
      \While {not converged}
        \State $X, Y \leftarrow$ mini-batch of samples from the training set
        \State $\bar{X}, Z \leftarrow$ AUTOENCODE($X, Y$)
        \State $\mathcal{L}_{cdvae}  \leftarrow \mathcal{L}_{cdvae}(\bar{X}, Y)$
        \State $\mathcal{L}_{wgan} \leftarrow \mathcal{L}_{wgan}(X)$
        \State $\mathcal{L}_{cls}  \leftarrow \mathcal{L}_{cls}(X, Y)$     
        \State
        
        \State // Update the discriminator and classifier
        \While {not converged}
          \State $\psi ~ \xleftarrow{update} ~ -\nabla_\psi (-\mathcal{L}_{wgan})$
          \State $\Psi ~ \xleftarrow{update} ~ -\nabla_\Psi \mathcal{L}_{cls}$
        \EndWhile
        \State
        
		\State // Update the encoder and generator
		\While {not converged}
          \State $\theta ~ \xleftarrow{update} ~ -\nabla_\theta (\mathcal{L}_{cdvae} + \alpha \mathcal{L}_{wgan} - \lambda \mathcal{L}_{cls})$
          \State $\phi ~ \xleftarrow{update} ~ -\nabla_\phi (\mathcal{L}_{cdvae} + \alpha \mathcal{L}_{wgan})$
        \EndWhile
        
      \EndWhile
      \end{algorithmic}
    \label{algo1}
    \end{algorithm}

\begin{table}[t]
\caption{Model architectures. Conv-h$\times$w-n indicates a convolutional layer with kernel size h$\times$w and n output channels. LReLU indicates the leaky ReLU activation function. FC indicates fully-connected linear layer . LN indicates the layer normalization layer.}
\label{tab:arch}
\centering
\begin{tabular}{c l}
\toprule

ConvLReLU & Conv-3x1-n, LN, LReLU \\
\midrule
ConvGLU & (Conv-3x1-n, LN, sigmoid) $\odot$ (Conv-3x1-n, LN, tanh) \\

\midrule
\midrule

$E_{SP}$ & \makecell[l]{ ConvLReLU $\times$ 5  (n=1024, 512, 256, 64, 32), \\
 FC-16 ($\mu$), FC-16 ($\sigma$)} \\

\midrule

$E_{MCC}$ & \makecell[l]{ ConvLReLU $\times$ 5  (n=512, 256, 128, 64, 32), \\
 FC-16 ($\mu$), FC-16 ($\sigma$)} \\

\midrule

$G_{SP}$ & \makecell[l]{ (Concat with \textbf{y}, ConvGLU) $\times$ 4 (n=128, 256, 512, 1024), \\
Concat with \textbf{y}, Conv-3x1-513} \\
\midrule
$G_{MCC}$ & \makecell[l]{ (Concat with \textbf{y}, ConvGLU) $\times$ 4 (n=64, 128, 256, 512), \\
Concat with \textbf{y}, Conv-3x1-513} \\

\midrule

$D_{SP}$ & \makecell[l]{ ConvLReLU $\times$ 5  (n=1024, 512, 256, 64, 32), FC-1 }\\
\midrule
$D_{SP}$ & \makecell[l]{ ConvLReLU $\times$ 5  (n=512, 256, 128, 64, 32), FC-1 }\\

\midrule

$C$ & \makecell[l]{ (Concat with \textbf{y}, ConvGLU) $\times$ 4 (n=128, 256, 512, 1024), \\
Concat with \textbf{y}, Conv-3x1-513} \\

\bottomrule

\end{tabular}
\end{table}

\section{Experimental evaluations}
\label{sec:experiments}

\subsection{Experimental settings}
\label{ssec:settings}

We conducted all experiments on the Voice Conversion Challenge (VCC) 2018 dataset, which contained recordings of 12 professional US English speakers with a sampling rate of 22050 Hz. The training and testing sets, respectively, consisted of 81 utterances and 35 utterances per speaker. We further divided the training utterances into 70/11 training/validation sets. The WORLD vocoder was used to extract acoustic features, including 513-dimensional SPs, 513-dimensional aperiodicity signals (APs), and fundamental frequency ($F_0$). 35-dimensional MCCs were then extracted from the SPs, which were then normalized to unit-sum, and the normalizing factor was used as the energy of SPs. The 0-th coefficient of MCCs was taken out as the energy of MCCs. We further applied Min-Max normalization to SPs and MCCs. In the conversion phase, the converted SPs in VAE systems and the converted MCCs in CDVAE systems (excluding CDVAE-GAN with $D_{\scaleto{SP}{3pt}}$) were obtained. The energy and AP were kept unmodified, and $F_0$ was converted using a linear mean-variance transformation in the log-$F_0$ domain.

The detailed network architectures are shown in Table~\ref{tab:arch}. We adopted the fully convolutional network (FCN) \cite{FCN} based CDVAE-VC as our baseline system \cite{F0-FCN-CDVAE}, which consumes continuous spectral frames extracted from the whole utterance and outputs a sequence of converted frames of the same length. This model has been confirmed to outperform the frame-wise CDVAE-VC counterpart. We also adopted a gradient penalty regularization \cite{WGAN-GP} in the WGAN objective to stabilize the training. Layer normalization \cite{layernorm}, the gated linear units activation function, and skip connections were also used to more effectively propagate the conditional information.

Following \cite{F0-FCN-CDVAE}, the latent space and speaker representation were set to 16-dimensional. We used a mini-batch of 16 and the Adam optimizer with a fixed learning rate of 0.0001. The hyper-parameters $\alpha$ and $\lambda$ were set to be 50 and 1000, respectively, according to a held-out validation set. For CDVAE-GAN, we first pre-trained the CDVAE for 100000 steps. Then, we adversarially trained the discriminator(s) with the whole network for 10000 steps. We followed a common WGAN training scheme \cite{WGAN, WGAN-GP} such that the discriminator(s) were updated for 5 iterations followed by 1 iteration of encoder and decoder update. For CDVAE-CLS-GAN, after training the CDVAE for 100000 steps, we pre-trained the classifier with the latent code extracted from the encoders for 30000 steps. Then, we trained the whole network for 10000 steps. After experimenting with different training schemes, here we updated the discriminator and the classifier for 1 iteration followed by 5 iterations of encoder and decoder update.

The following models are compared in order to examine the effectiveness of our proposed methods.

\begin{itemize}
    \item \textbf{VAE}: The FCN version of the VAE-VC model introduced in \cite{VAE-VC}. This model is only used to evaluate the impact of cross domain features on the degree of disentanglement.
    \item \CDVAE: The FCN model in \cite{F0-FCN-CDVAE}, which is the baseline model in our experiments.
    \item \CGS: The CDVAE with $D_{\scaleto{SP}{3pt}}$.
    \item \CGM: The CDVAE with $D_{\scaleto{MCC}{3pt}}$.
    \item \CGB: The CDVAE with $D_{\scaleto{SP}{3pt}}$ and $D_{\scaleto{MCC}{3pt}}$.
    \item \CL: The CDVAE  with CLS.
    \item \CLGS: The CDVAE  with $D_{\scaleto{SP}{3pt}}$ and CLS.
    \item \CLGM: The CDVAE with $D_{\scaleto{MCC}{3pt}}$ and CLS.
\end{itemize}

For simpilcity, in the rest of the paper, we use brackets to surround the type of feature used during conversion, and that path will be used in CDVAE-based methods. For instance, \CGM\ [MCC] uses the MCC and the MCC-MCC path. In addition, if MCC is used in \CDVAE\ and \CL, we additionally compare systems incorporating the global variance (GV) post-filter \cite{fastGV} to enhance the output, as in the original CDVAE \cite{CDVAE}.

\subsection{Evaluation methodology}
\label{ssec:metrics}

\subsubsection{Objective evaluation metrics}
\begin{itemize}
	\item Mel-Cepstrum distortion (MCD): MCD measures the spectral distortion in the MCC domain, and is a commonly adopted objective metric in the field of VC. It is calculated as:
	\begin{equation}
	    MCD [dB] = \frac{10}{\log 10} \sqrt{2 \sum_{d=1}^K (mcc^{(c)}_d - mcc^{(t)}_d )^2} ,
	\end{equation}
	where $K$ is the dimension of the MCCs and $mcc^{(c)}_d$ and $mcc^{(t)}_d$ represent the $d$-th dimensional coefficient of the converted MCCs and the target MCCs, respectively. In practice, MCD is calculate in a utterance-wise manner. A dynamic time warping (DTW) based alignment is performed to find the corresponding frame pairs between the non-silent converted and target MCC sequences beforehand. 
	
	\item Global variance (GV): GV serves as a metric for the over-smoothness of the output features. GV is usually calculated dimension-wise over all non-silent frames in the evaluation set. The $d$-dimensional GV value is calculated as follows:
	\begin{equation}
	    GV[d] = \frac{1}{N} \sum_{n=1}^N (mcc^{(c)}_{n,d} - \overline{mcc}^{(c)}_{d})^2 ,
	\end{equation}
	where $\overline{mcc}^{(c)}_d$ is the mean of all converted $d$-th dimensional MCC coefficients.
	
	\item Modulation Spectrum (MS): MS \cite{MS} is defined as the log-scaled power spectrum of a given feature sequence. The temporal fluctuation of the sequence is first decomposed into individual modulation frequency components, and their power values are represented as the MS. In this work we measure the MS of MCCs. Different from previous works that measured the MS of specific dimension of the MCC sequence, here we report the average of all dimensions. We also measure a MS distortion (MSD), where the MSD for the $d$-dimension is calculated by:
	\begin{equation}
	    MSD[d] = \sqrt{\frac{1}{N} \sum_{n=1}^N (mcc^{(t)}_{n,d} - mcc^{(c)}_{n,d})^2}.
	\end{equation}
	
\end{itemize}

\subsubsection{Subjective evaluation methods}
We recruited 14 participants for the following two subjective evaluations.\footnote{A demo web page with samples used for subjective evaluation is available at \url{ https://unilight.github.io/CDVAE-GAN-CLS-Demo/}.}
\begin{itemize}
    %\item AB test on naturalness: In this test, each participant was demanded to choose the one with more natural voice among two converted utterances generated by the two methods for the same sentence (content) in random order.
    \item The mean opinion score (MOS) test on naturalness: Subjects were asked to evaluate the naturalness of the converted and natural speech samples on a scale from 1 (completely unnatural) to 5 (completely natural).
    \item The VCC \cite{vcc2018} style test on similarity: This paradigm was adopted by the VCC organizing committee. Listeners were given a pair of speech utterances consisting of a natural speech sample from a target speaker and a converted speech sample. Then, they were asked to determine whether the pair of utterances can be produced by the same speaker, with a 4-level confidence of their decision, i.e., sure or not sure.
\end{itemize}

\begin{table}[t]
\renewcommand{\arraystretch}{1.3}
\caption{Mean Mel-cepstral distortions [dB] of all non-silent frames in the evaluation set for the compared models.}
\label{tab:MCD}
\centering
\begin{tabular}{l c c c c c}
\toprule
Model & F-F & M-M & M-F & F-M & Avg.\\
\midrule
CDVAE [MCC] & 6.56 & 5.76 & 6.96 & 6.27 & 6.39\\
\midrule
CDVAE-GAN\textsubscript{SP} [SP] & 7.09 & 6.38 & 7.38 & 6.93 & 6.94\\
CDVAE-GAN\textsubscript{MCC} [MCC] & 7.52 & 6.65 & 7.87 & 7.27 & 7.33\\
CDVAE-GAN\textsubscript{BOTH} [MCC] & 7.44 & 6.85 & 7.90 & 7.30 & 7.37\\
\midrule
CDVAE-CLS [MCC] & 6.65 & 6.29 & 7.02 & 6.49 & 6.61\\
\midrule
CDVAE-CLS-GAN\textsubscript{SP} [SP]& 7.23 & 6.91 & 7.64 & 7.08 & 7.21\\
CDVAE-CLS-GAN\textsubscript{MCC} [MCC]& 7.71 & 6.62 & 7.76 & 7.05 & 7.29\\
CDVAE-CLS-GAN\textsubscript{BOTH} [MCC] & 7.57 & 7.13 & 8.12 & 7.30 & 7.53\\

\hline
\end{tabular}
\end{table}

\begin{figure*}[t]
	\centering
	
	\begin{subfigure}[b]{0.32\textwidth}
		\centering
		\includegraphics[width=\textwidth]{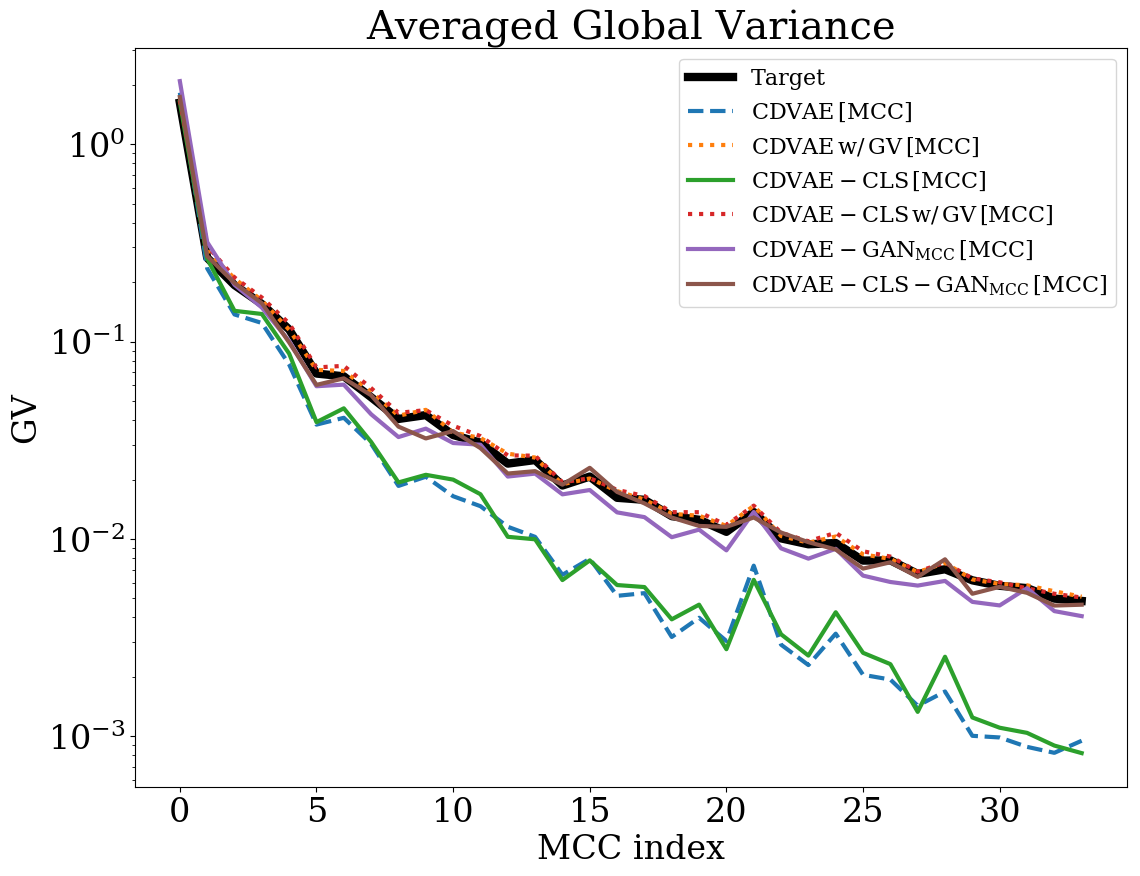} 
		\caption{\label{fig:gv-all}}
	\end{subfigure}
	\begin{subfigure}[b]{0.32\textwidth}
		\centering
		\includegraphics[width=\textwidth]{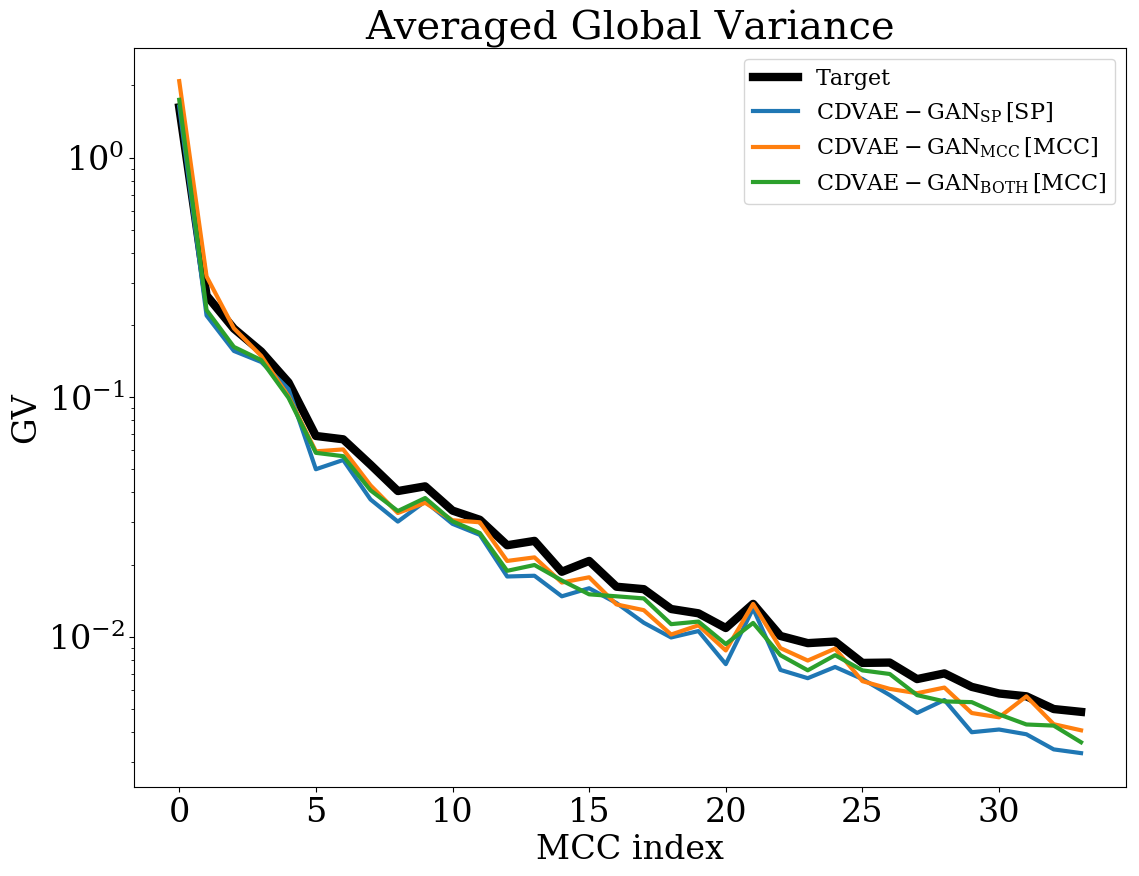} 
		\caption{\label{fig:gv-gan}}
	\end{subfigure}
	\begin{subfigure}[b]{0.32\textwidth}
		\centering
		\includegraphics[width=\textwidth]{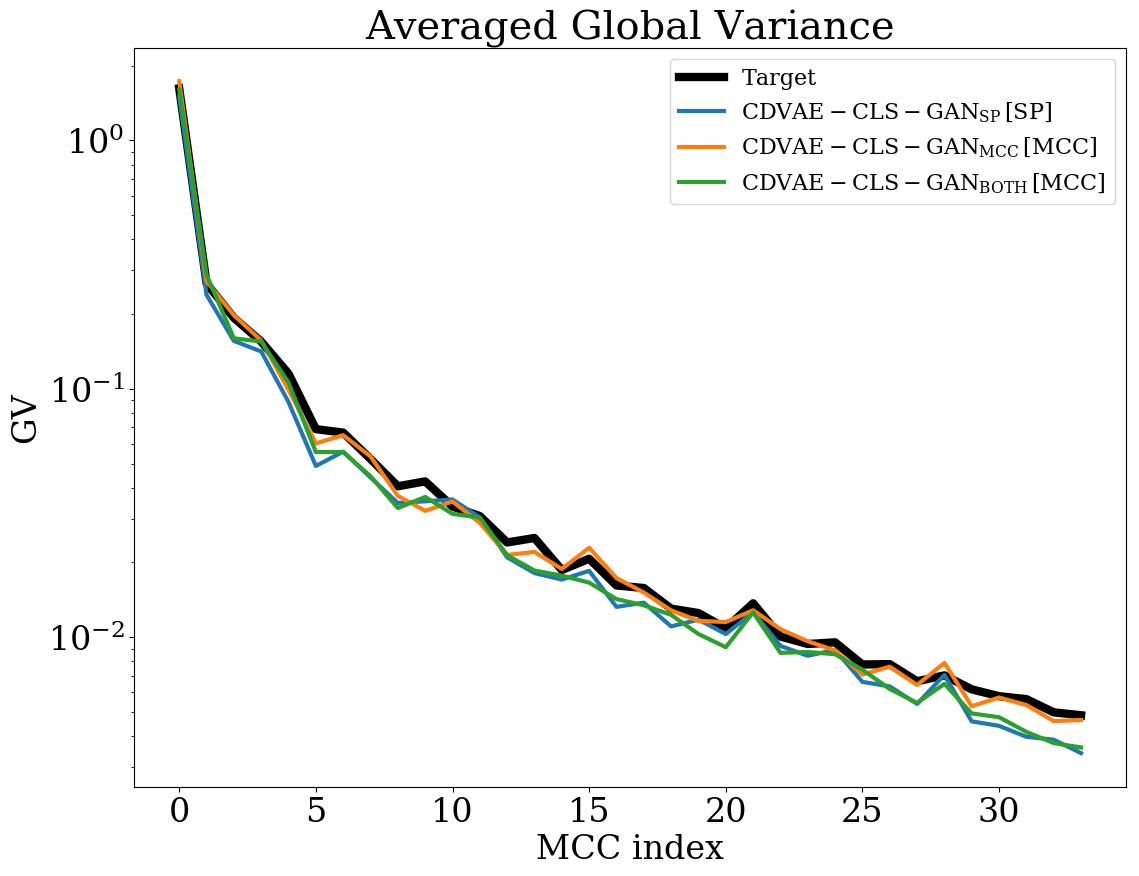} 
		\caption{\label{fig:gv-cls-gan}}
	\end{subfigure}
	
	\caption{Global variance curves of all non-silent frames averaged over all conversion pairs for the compared models.	\label{fig:gv}}
\end{figure*}

\begin{figure*}[t]
	\centering
	
	\begin{subfigure}[b]{0.32\textwidth}
		\centering
		\includegraphics[width=\textwidth]{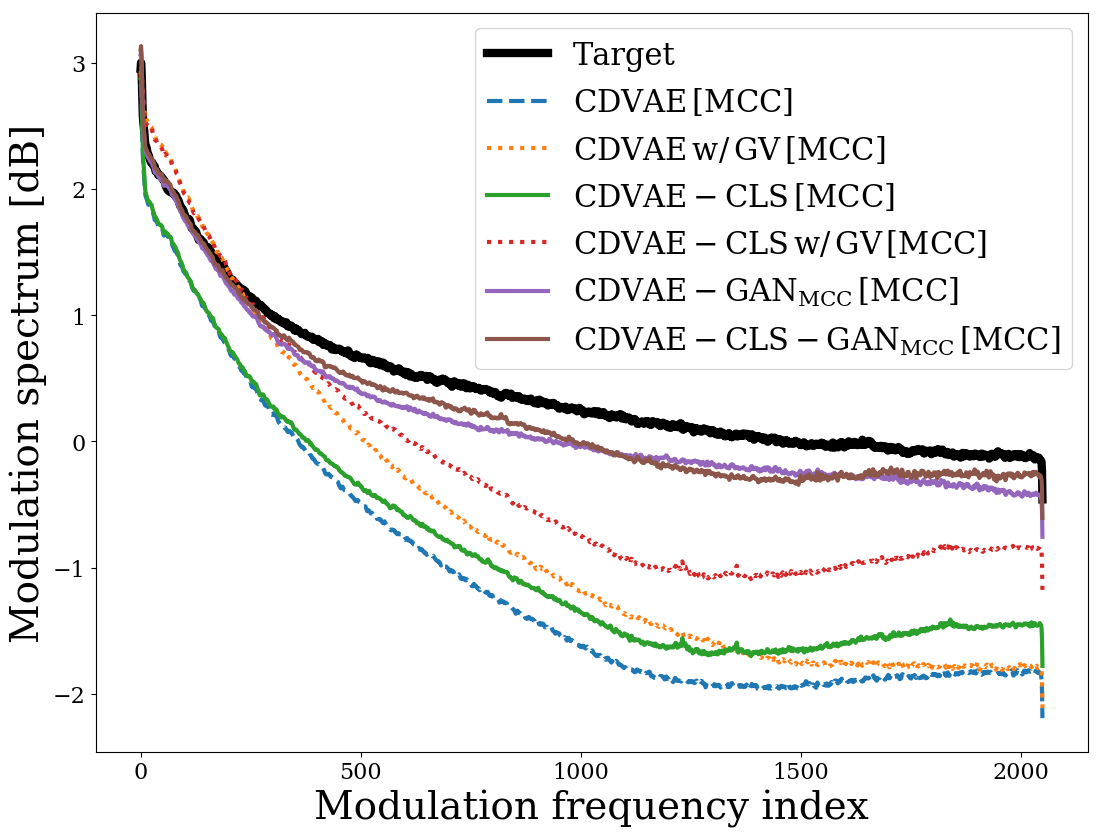} 
		\caption{\label{fig:ms-all}}
	\end{subfigure}
	\begin{subfigure}[b]{0.32\textwidth}
		\centering
		\includegraphics[width=\textwidth]{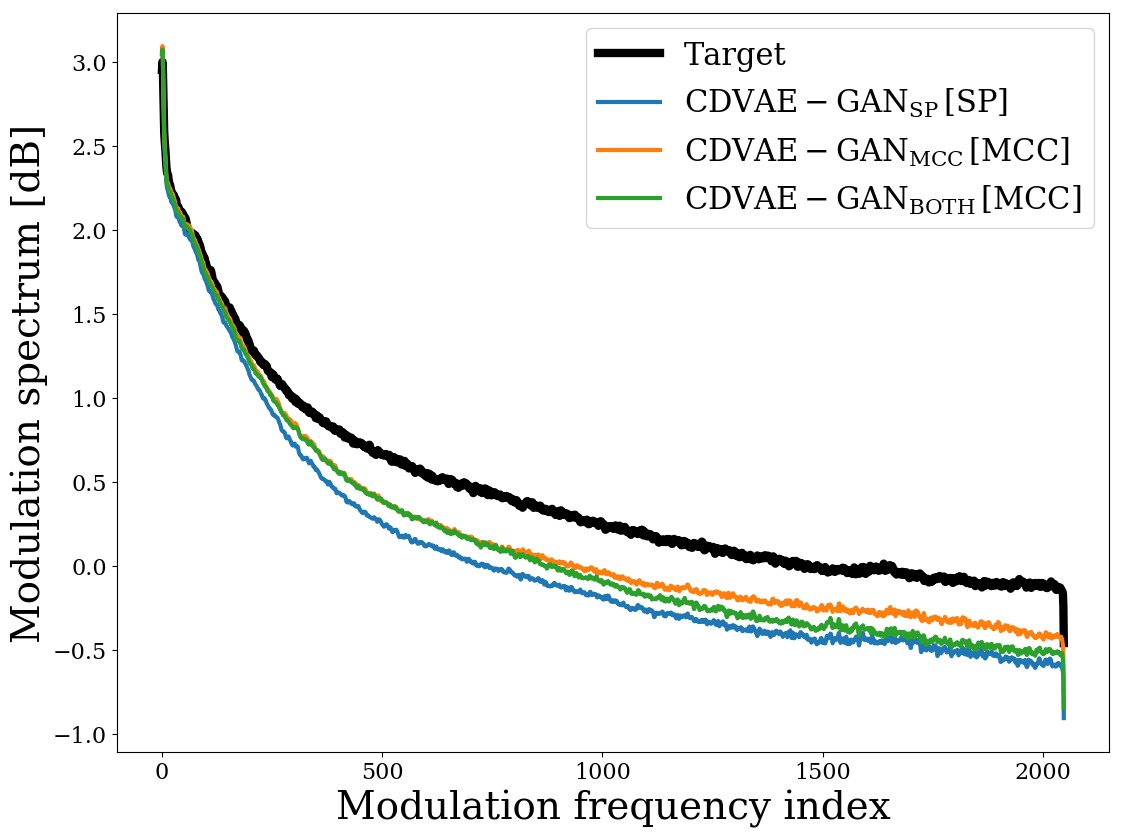} 
		\caption{\label{fig:ms-gan}}
	\end{subfigure}
	\begin{subfigure}[b]{0.32\textwidth}
		\centering
		\includegraphics[width=\textwidth]{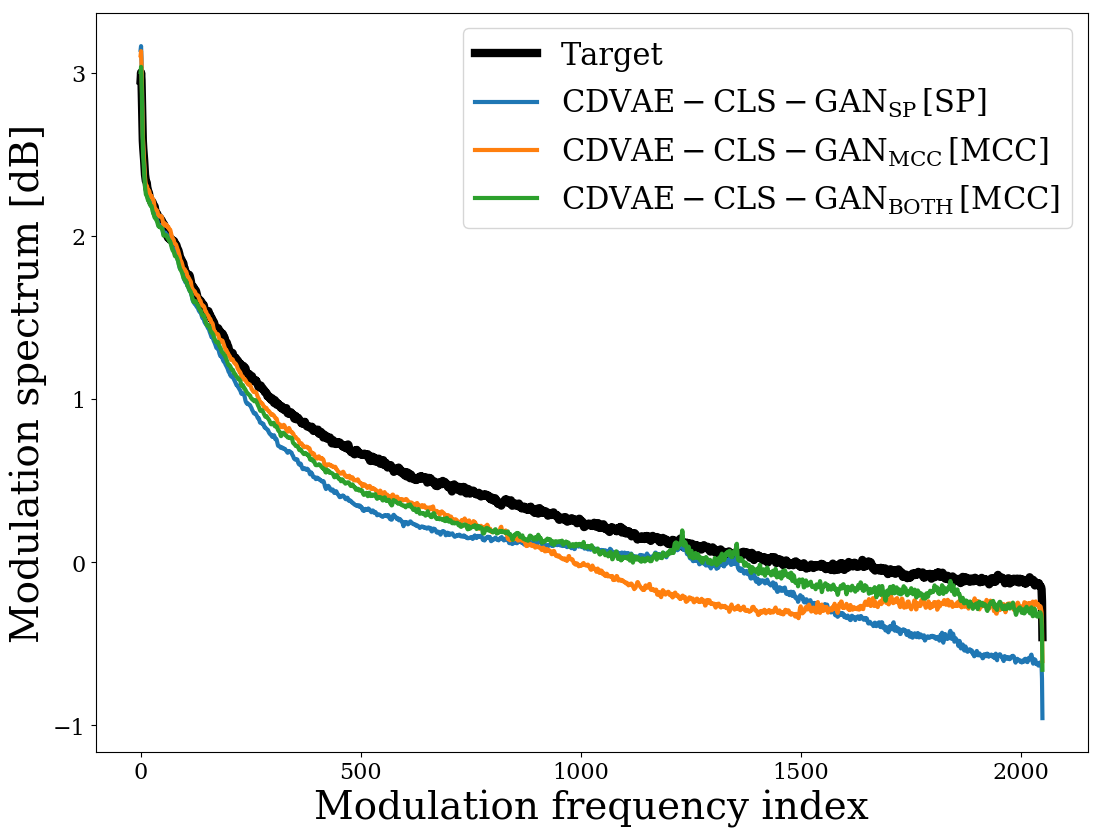} 
		\caption{\label{fig:ms-cls-gan}}
	\end{subfigure}
	
	\caption{Average modulation spectrum curves over all dimensions of all non-silent frames over all conversion pairs for the compared models.	\label{fig:MS}}
\end{figure*}

\subsection{Applying GANs to different features}
\label{ssec:exp-gan-features}

We first compare \CGS, \CGM, \CGB\ and \CLGS, \CLGM, \CLGB, respectively. As in Table~\ref{tab:MCD}, \CGB\ and \CLGB\ gave the highest MCD, while in Figures~\ref{fig:gv-gan}, \ref{fig:gv-cls-gan}, \ref{fig:ms-gan} and \ref{fig:ms-cls-gan}, we can see that in terms of GV and MS, \CGM\ and \CLGM\ yielded curves closer to the target curves, where the curves of the other models deviated more from the target curves. Meanwhile, consistent with a common observation in the VC literature that MCD, which measures the sample mean, often yields opposite results to GV and MS, both presenting the sample variance \cite{GMM-VC, GAN-PF-SPSS}. This result suggests that modeling both feature domains simultaneously does not always yield better results. As for perceptual performance, our internal listening tests revealed that \CGM\ gave the best results among the three models. Note that although \CGS\ and \CLGS\ gave the lowest MCD compared with the other two models, they do not necessarily outperform their MCC counterparts in listening tests. We speculate that fitting the SP domain tends to give more over-smoothed output features, resulting in low MCDs but not beneficial for improving perceptual performance. The result is reasonable since the MCC-MCC path is used when performing conversion.

\subsection{Effectiveness of GANs}
\label{ssec:exp-gan}

\begin{figure}[t]
  \centering
  \includegraphics[width=0.48\textwidth]{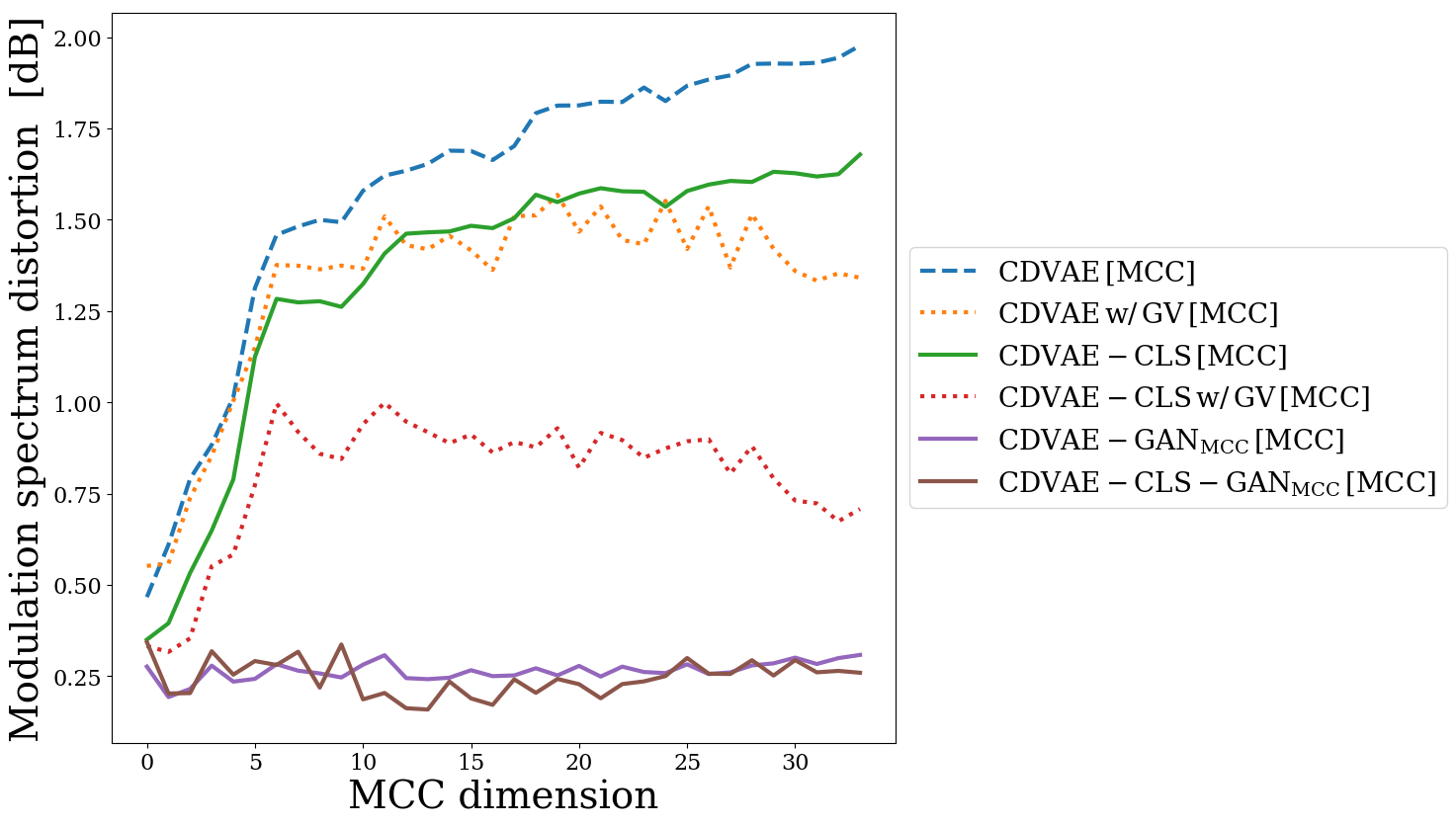}
  \centering
  \caption{Modulation spectrum distortion curves of all non-silent frames over all conversion pairs for the compared models.
  }
  \label{fig:msd}
\end{figure}

Next, we examine the effectiveness of combining GANs with \CDVAE\ and \CL. Based on the discussion in the previous subsection, we focus on \CGM\ and \CLGM\ here. As in Figures~\ref{fig:gv-all}, \ref{fig:ms-all} and \ref{fig:msd}, we can see that \CGM\ and \CLGM\ fit the GV and MS statistics to the target much better than \CDVAE\ and \CL, respectively. Also, models with GAN yield very small MSD comparing to the rest of the models. This confirms that involving a GAN objective in training indeed improves the modeling of the statistics, especially the variance of real speech data. Table~\ref{tab:mos} and Figure~\ref{fig:similarity} show the subjective evaluation results. The $t$-test showed that \CGM\ significantly outperformed \CDVAE\ with a $p$-value of $4.21\times 10^{-5}$. Meanwhile, \CLGM\ significantly outperformed \CL\ with a $p$-value of $4.02\times 10^{-4}$. These results confirm the effectiveness of GAN. On the other hand, \CGM\ performed comparably with \CDVAE\ with GV post-processing ($p$-value = $4.03\times 10^{-2}$), while \CLGM\ performed comparably with \CL\ with GV post-processing ($p$-value = $7.64\times 10^{-2}$). These results are consistent with our findings in the objective evaluations, suggesting that GANs enhance the variance of output features, thus have the potential to replace the GV post-filtering process commonly involved in traditional MCC-based VC systems \cite{GMM-VC}. This is advantageous since the model can then be freed from the post-filtering process in the online conversion phase, which may benefit real-time applications.

\begin{table*}[t]
\renewcommand{\arraystretch}{1.3}
\caption{Mean opinion scores on naturalness for the compared models and the natural target voice with 95\% confidence intervals.}
\label{tab:mos}
\centering
\begin{tabular}{l c c c c c}
\toprule
Model & F-F & M-M & M-F & F-M & Avg.\\
\midrule
CDVAE [MCC] & 2.50 $\pm$ 0.27 & 2.42 $\pm$ 0.21 & 2.31 $\pm$ 0.25 & 2.28 $\pm$ 0.31 & 2.40 $\pm$ 0.28 \\
CDVAE-CLS [MCC] & 2.72 $\pm$ 0.37 & 2.61 $\pm$ 0.48 & 2.44 $\pm$ 0.30 & 2.17 $\pm$ 0.29 & 2.55 $\pm$ 0.39 \\
CDVAE w/ GV [MCC] & 3.36 $\pm$ 0.44 & 3.36 $\pm$ 0.26 & 2.94 $\pm$ 0.30 & 2.89 $\pm$ 0.50 & 3.12 $\pm$ 0.30 \\
CDVAE-CLS w/ GV [MCC] & 3.50 $\pm$ 0.53 & 3.53 $\pm$ 0.49 & 3.06 $\pm$ 0.31 & 3.17 $\pm$ 0.38 & 3.30 $\pm$ 0.38 \\
CDVAE-GAN\textsubscript{MCC} [MCC] & 3.19 $\pm$ 0.40 & 3.06 $\pm$ 0.38 & 2.61 $\pm$ 0.36 & 2.58 $\pm$ 0.27 & 2.95 $\pm$ 0.30 \\
CDVAE-CLS-GAN\textsubscript{MCC} [MCC] & 2.94 $\pm$ 0.37 & 3.58 $\pm$ 0.48 & 3.06 $\pm$ 0.26 & 3.11 $\pm$ 0.57 & 3.15 $\pm$ 0.36 \\
\midrule
Target & - & - & - & - & 4.75 $\pm$ 0.25\\
\bottomrule
\end{tabular}
\end{table*}

\subsection{Effectiveness of CLS}
\label{ssec:exp-cls}

Next, we evaluate the effectiveness of the adversarial speaker classifier. Looking at the \CDVAE, \CG\ models and their counterparts with CLS, a trend of increase in MCD values can be observed in Table~\ref{tab:MCD}. On the other hand, Figures~\ref{fig:gv-all}, \ref{fig:ms-all} and \ref{fig:msd} show that applying CLS to \CDVAE\ and \CLGM\ yields similar GV values, but with MS values closer to those of the target, as well as a smaller MSD. These results imply that CLS can improve objective statistics.

\begin{figure}[t]
  \centering
  \includegraphics[width=0.48\textwidth]{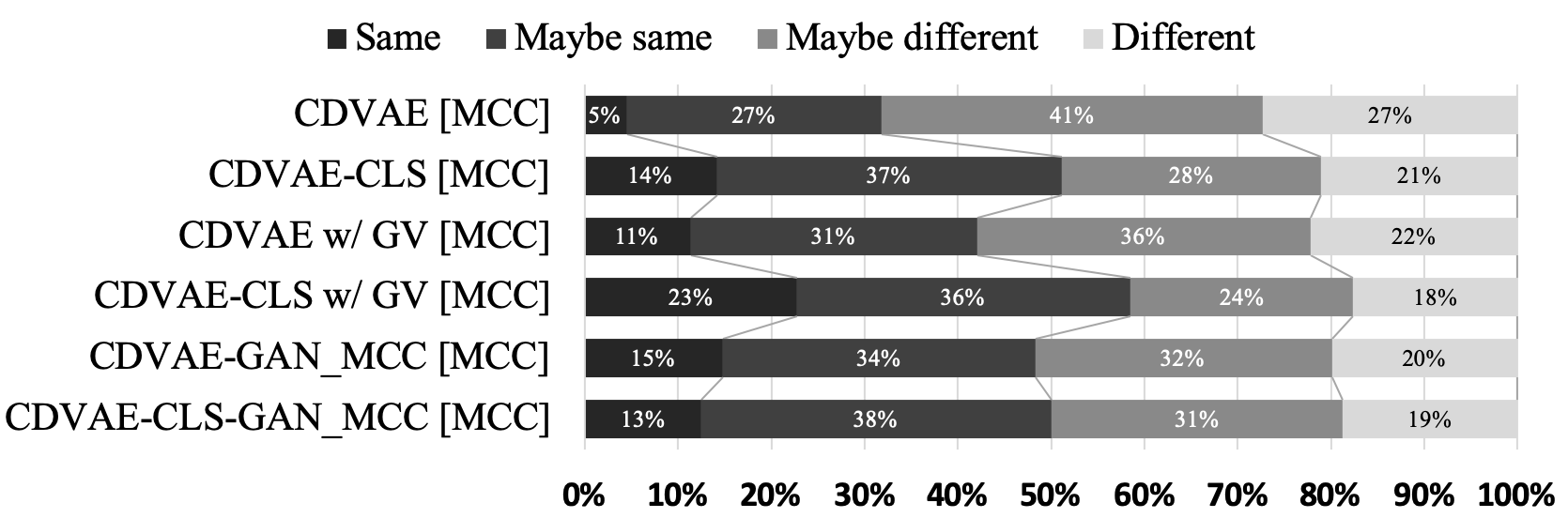}
  \centering
  \caption{Similarity results over all speaker pairs for the compared models.}
  \label{fig:similarity}
\end{figure}

Table~\ref{tab:mos} and Figure~\ref{fig:similarity} show the subjective evaluation results. The effectiveness of CLS can be confirmed by the following observations: The speech naturalness was improved in all conversion pairs, by adding CLS to \CDVAE, \CDVAE\ w/ GV, and \CGM. This is consistent with our aforementioned findings from the objective evaluations. Furthermore, the conversion similarity is greatly improved when incorporating CLS in \CDVAE\ and \CDVAE\ w/ GV, and is slightly improved when added to \CGM. This confirms our initial motivation of CLS, which is to increase speaker similarity by eliminating source speaker identity in the latent code.

%\begin{figure}[t]
%  \centering
%  \includegraphics[width=0.5\textwidth]{mos-revised.png}
%  \centering
%  \caption{Mean opinion scores on naturalness over all pairs for the compared models and the natural target voice. Error bars indicate the 95\% confidence intervals.
%  }
%  \label{fig:mos}
%\end{figure}

\subsection{Disentanglement Measure}
\label{ssec:exp-dod}

In this section, we investigate the degree of disentanglement of the VC models involved in this study. We use a novel metric that was recently proposed in \cite{F0-FCN-CDVAE} as the disentanglement measurement, termed DEM. The main design concept of DEM is that a pair of sentences of the same content uttered by the source and target speakers should have similar latent codes since the phonetic contents are the same. Therefore, we can use the cosine similarity to measure the distance of the latent codes obtained from the paired utterances. Specifically, the procedure to calculate DEM is as follows:
\begin{enumerate}
    \item extracting the latent codes of a pair of parallel utterances spoken by the source and target speakers;
    \item aligning the frame sequences of the pair of utterances using DTW;
    \item calculating the frame-wise cosine similarity, and then taking the average of the entire sequence. 
\end{enumerate}
As with other popular evaluation metrics, e.g., MCD and MSD, computing DEM requires parallel data. Since parallel data are usually available in standardized VC datasets, DEM is a simple but effective measure of the degree of disentanglement of the latent codes.

\begin{table}[t]
\renewcommand{\arraystretch}{1.3}
\caption{The results of DEM: the cosine similarity of the latent codes extracted from non-silent frames of parallel utterances of source-target pairs.}
\label{tab:dod}
\centering
\begin{tabular}{l c c c c c}
\toprule
Model & F-F & M-M & M-F & F-M & Avg.\\
\midrule
VAE [SP] & .568 & .633 & .534 & .552 & .571\\
CDVAE [SP] & .597 & .658 & .557 & .577 & .597\\
CDVAE-GAN\textsubscript{SP} [SP] & .605 & .677 & .565 & .594 & \textbf{.610}\\
CDVAE-CLS [SP] & .573 & .582 & .508 & .535 & .550\\
CDVAE-CLS-GAN\textsubscript{SP} [SP] & .629 & .638 & .573 & .602 & \textbf{.610}\\
\midrule
CDVAE [MCC] & .530 & .588 & .476 & .502 & .524\\
CDVAE-GAN\textsubscript{MCC} [MCC] & .559 & .609 & .502 & .534 & .551\\
CDVAE-CLS [MCC] & .575 & .581 & .507 & .533 & .549\\
CDVAE-CLS-GAN\textsubscript{MCC} [MCC] & .583 & .621 & .561 & .563 & \textbf{.584}\\
\bottomrule
\end{tabular}
\end{table}

Table~\ref{tab:dod} shows the evaluation results of DEM. First, we observe that \CDVAESP\ yields higher DEM scores than \VAE\ [SP]. This confirms that introducing cross domain features indeed increases the degree of disentanglement. Next, comparing the corresponding methods in the upper and lower half of the table, which used SP and MCC as input features respectively, the DEM scores of the upper is consistently higher than those of the latter. This result is somehow reasonable because here SPs (513-dimensional) are of higher dimensions than MCCs (35-dimensional) and carry much detailed information. As a result, in terms of cosine similarity measure, higher DEM could be observed in the upper half methods than the lower half.

One interesting finding here is that when corporating GANs in \CDVAE\ and \CL\ models, the DEM scores are consistently and significantly improved. This result indicates that during training of \CGM, although not in our original expectaions, the discriminator not only benefits the decoders, but also indirectly guides the latent codes to be better disengagled.

As for CLS, we first observe that including CLS in \CDVAE\ improves the DEM score when using MCC yet degrades when using SP. Although this somewhat makes the effectiveness of CLS inconvincing, we note that \CL\ [SP] and \CL\ [MCC] have nearly identical DEM scores. This intersesting finding shows that the CLS forces the encoders to encode different features into similar contents. On the other hand, including CLS in \CG\ models boosts the DEM scores of cross gender pairs, which confirms that CLS can help the encoders eliminate speaker independent information, such as gender.

Finally, we compare the results of similarity tests of \CDVAEMCC, \CGM, and \CLGM\ in Figure~\ref{fig:similarity} and the DEM results in Table~\ref{tab:dod}. \CLGM\ achieves the highest similarity scores in Figure~\ref{fig:similarity} and gives the highest DEM scores in Table~\ref{tab:dod}. The result verifies the positive correlation between the conversion performance and the degree of disentanglement of the latent codes.

\section{Conclusions}
\label{sec:conclusion}

In this paper, we have extended the cross-domain VAE based VC framework by integrating GANs and CLS into the training phase. The GAN objective was used to better approximate the distribution of real speech signals. The CLS, on the other hand, was applied to the latent code as an explicit constraint to eliminate speaker-dependent factors. Objective and subjective evaluations confirmed the effectiveness of the GAN and CLS objectives. We have also investigated the correlation between the degree of disentanglement and the conversion performance. A novel evaluation metric, DEM, that measures the degree of disentanglement in VC was derived. Experimental results confirmed a positive correlation between the degree of disentanglement and the conversion performance.

In the future, we will exploit more acoustic features in the \CDVAE\ system, including rawer features, such as the magnitude spectrum, and hand-crafted features, such as line-spectral pairs. An effective algorithm that can optimally determine the latent space dimension is also worthy of study. Finally, it is worthwhile to generalize this disentanglement framework to extract speaker-invariant latent representation from unknown source speakers in order to achieve many-to-one VC.

We have made the source code publicly accessible so that readers can reproduce our results.\footnote{\url{https://github.com/unilight/cdvae-vc}}

\section*{Acknowledgment}

This work was supported in part by the MOST-Taiwan Grants MOST 107-2221-E-001-008-MY3 and MOST 108-2634-F-001-004.



\bibliographystyle{IEEEtran}
\bibliography{ref}


\begin{IEEEbiography}[{\includegraphics[width=1in,height=1.25in,clip,keepaspectratio]{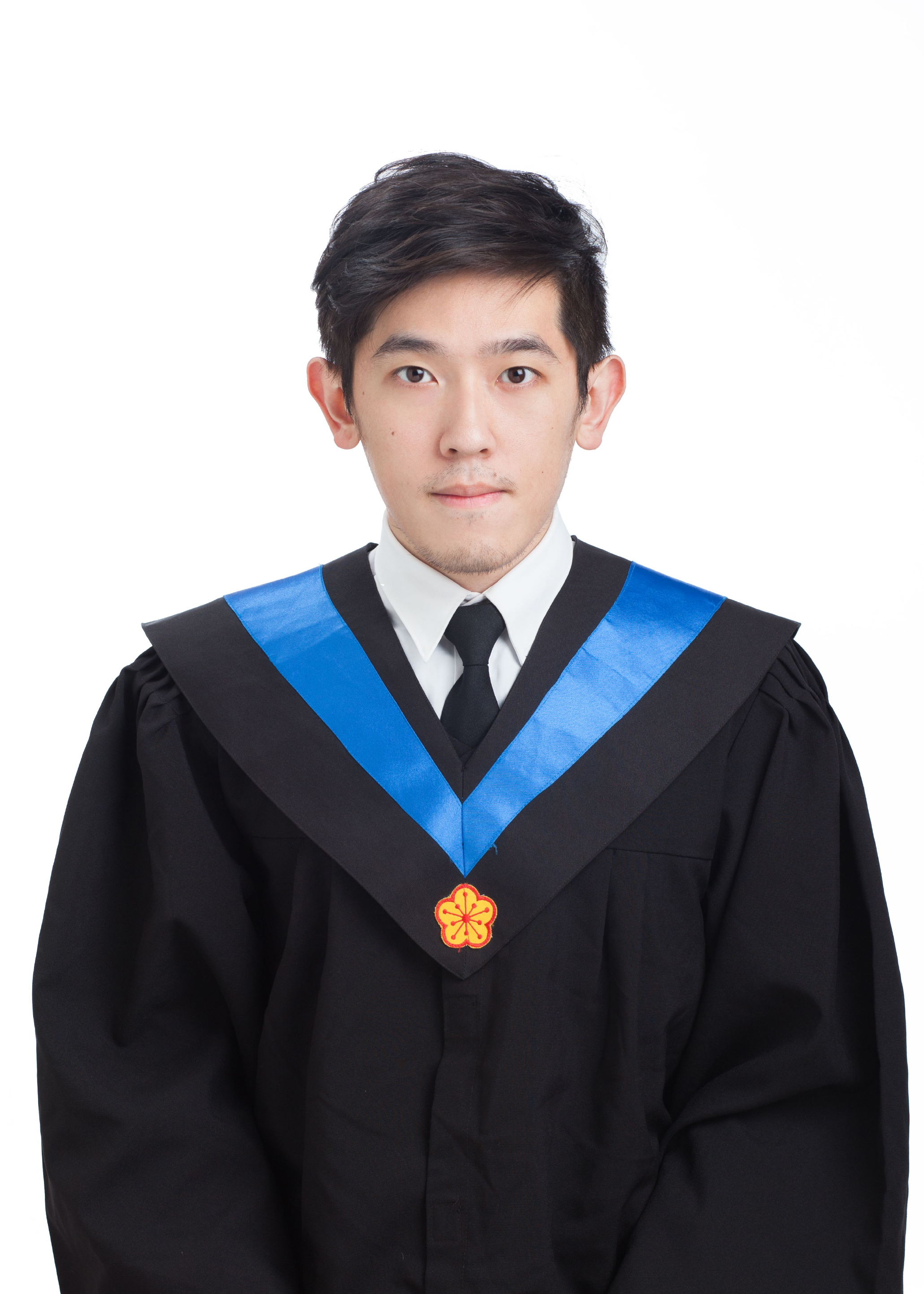}}]{Wen-Chin Huang} (S'20) received the B.S. degree in computer science and information engineering from National Taiwan University, Taipei, Taiwan, in 2018. He is currently working towards his M.S. degree in Nagoya University, Nagoya, Japan. He was a research assistant in the Institute of Information Science, Academia Sinica, Taipei, Taiwan, from 2017 to 2019. His research interests include voice conversion, speech synthesis and deep learning. He received the best student paper award in ISCSLP2018.
\end{IEEEbiography}

\begin{IEEEbiography}[{\includegraphics[width=1in,height=1.25in,clip,keepaspectratio]{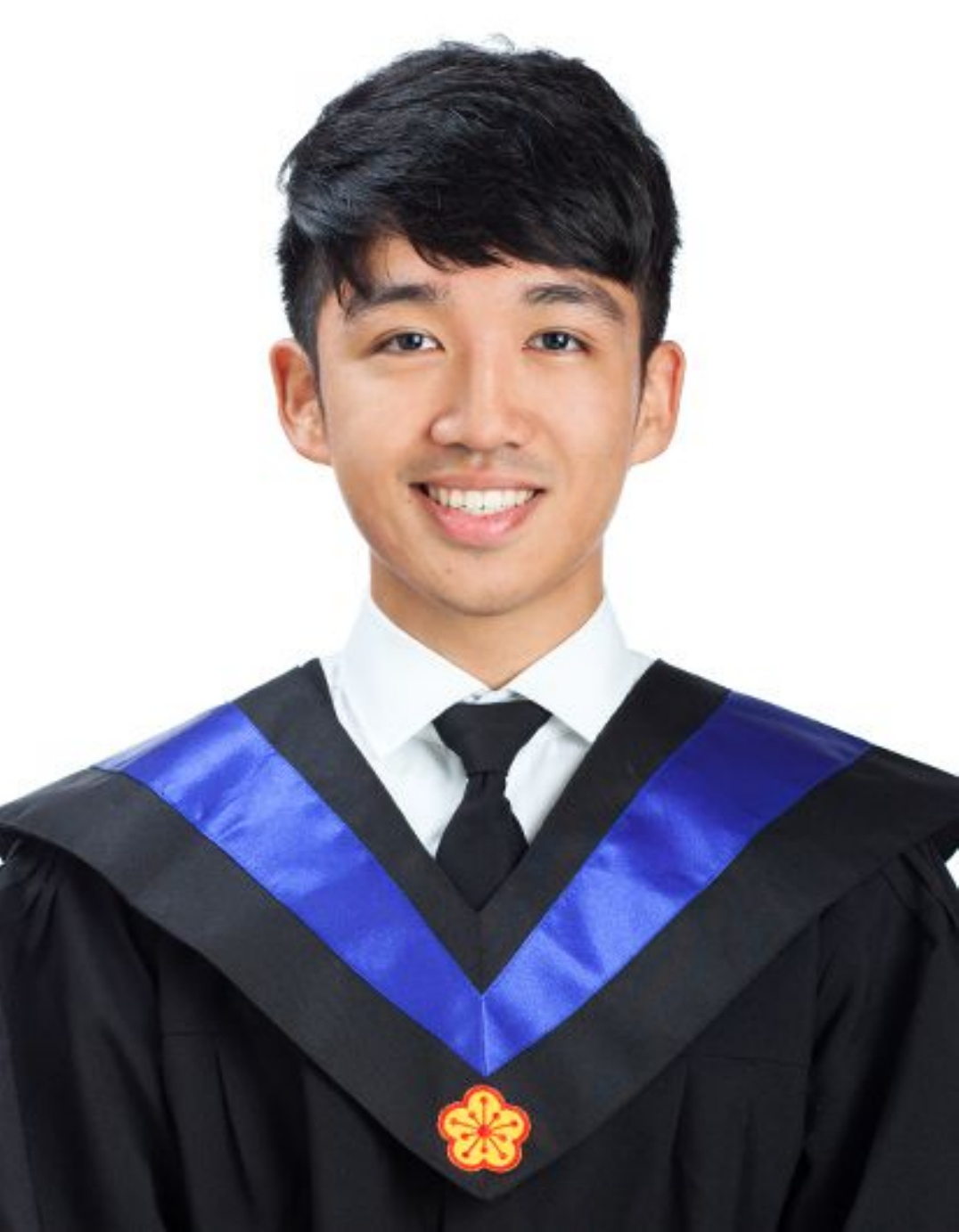}}]{Hao Luo} received the B.S. degree in electrical engineering from National Taiwan University, Taipei, Taiwan. He is currently pursuing the M.S. degree in electrical engineering from National Taiwan University, Taipei, Taiwan. He is an intern in Institute of Information Science, Academia Sinica, Taipei, Taiwan. His research interest is voice conversion.
\end{IEEEbiography}

\begin{IEEEbiography}[{\includegraphics[width=1in,height=1.25in,clip,keepaspectratio]{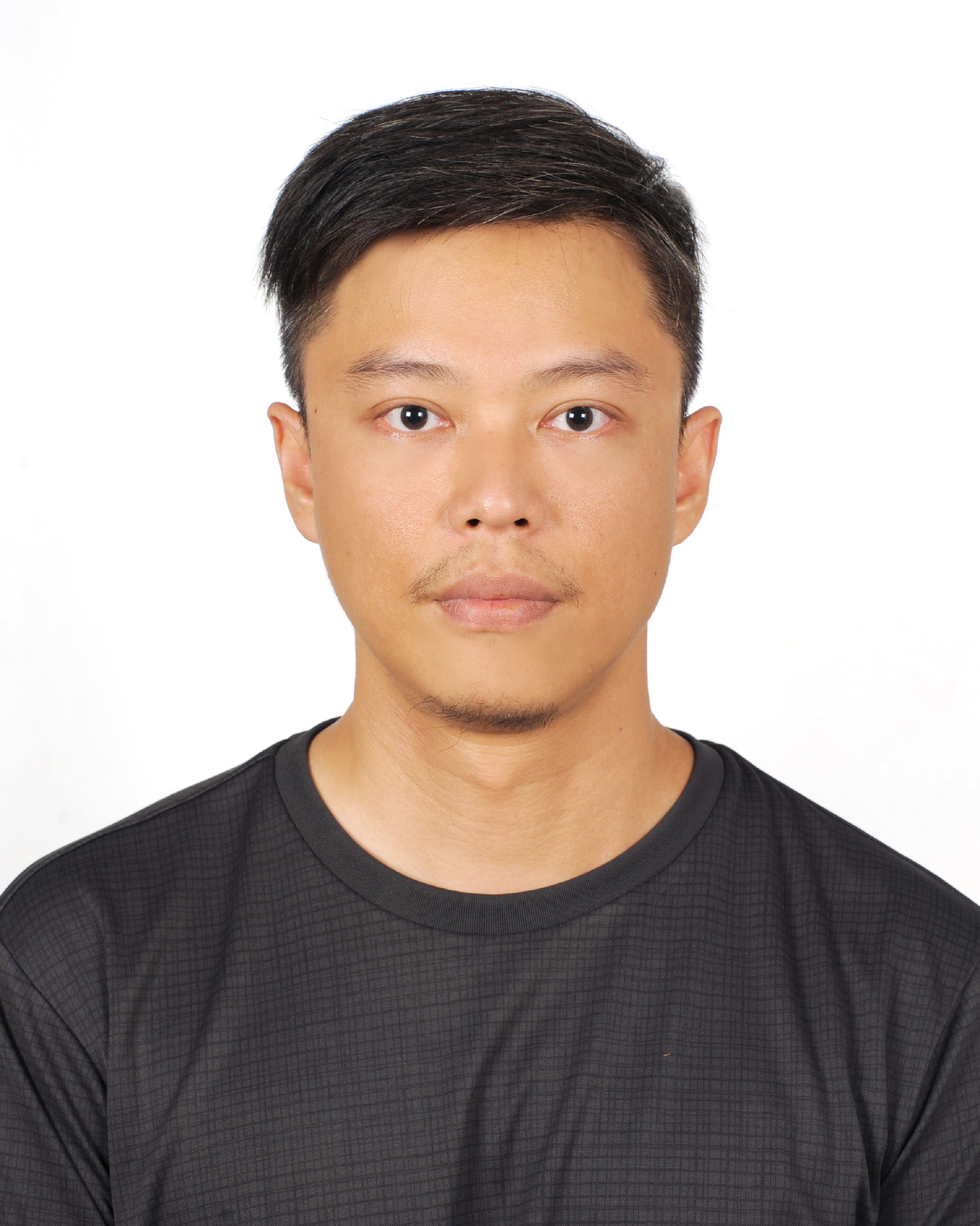}}]{Hsin-Te Hwang} received the Ph.D degree in Institute of Communications Engineering, National Chiao Tung University, Hsinchu, Taiwan, in 2017. From 2010 to 2017, he was a Research Assistant at the Institute of Information Science, Academia Sinica, Taipei, Taiwan. He was a Postdoctoral Fellow at the Institute of Information Science, Academia Sinica, Taipei, Taiwan, from 2017 to 2018. He joined the Intelligo Technology Inc. Taiwan Branch in 2019, where he is currently a Senior Engineer. His research interests include speech signal processing, particularly, voice conversion, speech enhancement, and speech synthesis.
\end{IEEEbiography}

\begin{IEEEbiography}[{\includegraphics[width=1in,height=1.25in,clip,keepaspectratio]{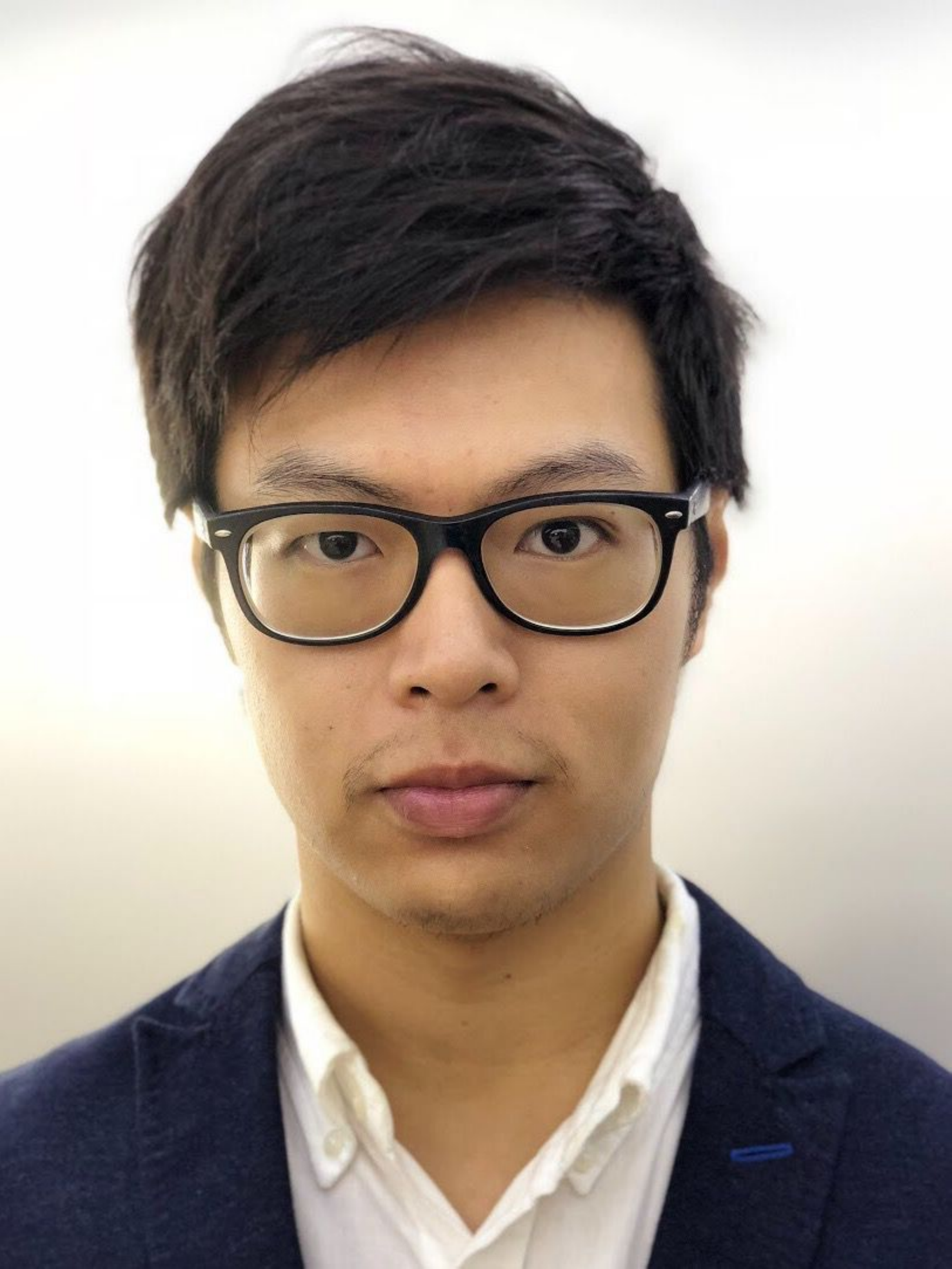}}]{Chen-Chou Lo} received the M.S. degree in electrical engineering from National Central University, Taoyuan, Taiwan. He is currently pursuing a Ph.D. degree in the Department of Electrical Engineering at Katholieke Universiteit Leuven, Belgium. From 2018 to 2019, he was a Research Assistant at the Institute of Information Science, Academia Sinica, Taipei, Taiwan. He joined the research group EAVISE at Katholieke Universiteit Leuven in October 2019, where he is currently a Ph.D. student. His research interests are in audiovisual signal processing, speech enhancement, and converted speech assessments.
\end{IEEEbiography}

\begin{IEEEbiography}[{\includegraphics[width=1in,height=1.25in,clip,keepaspectratio]{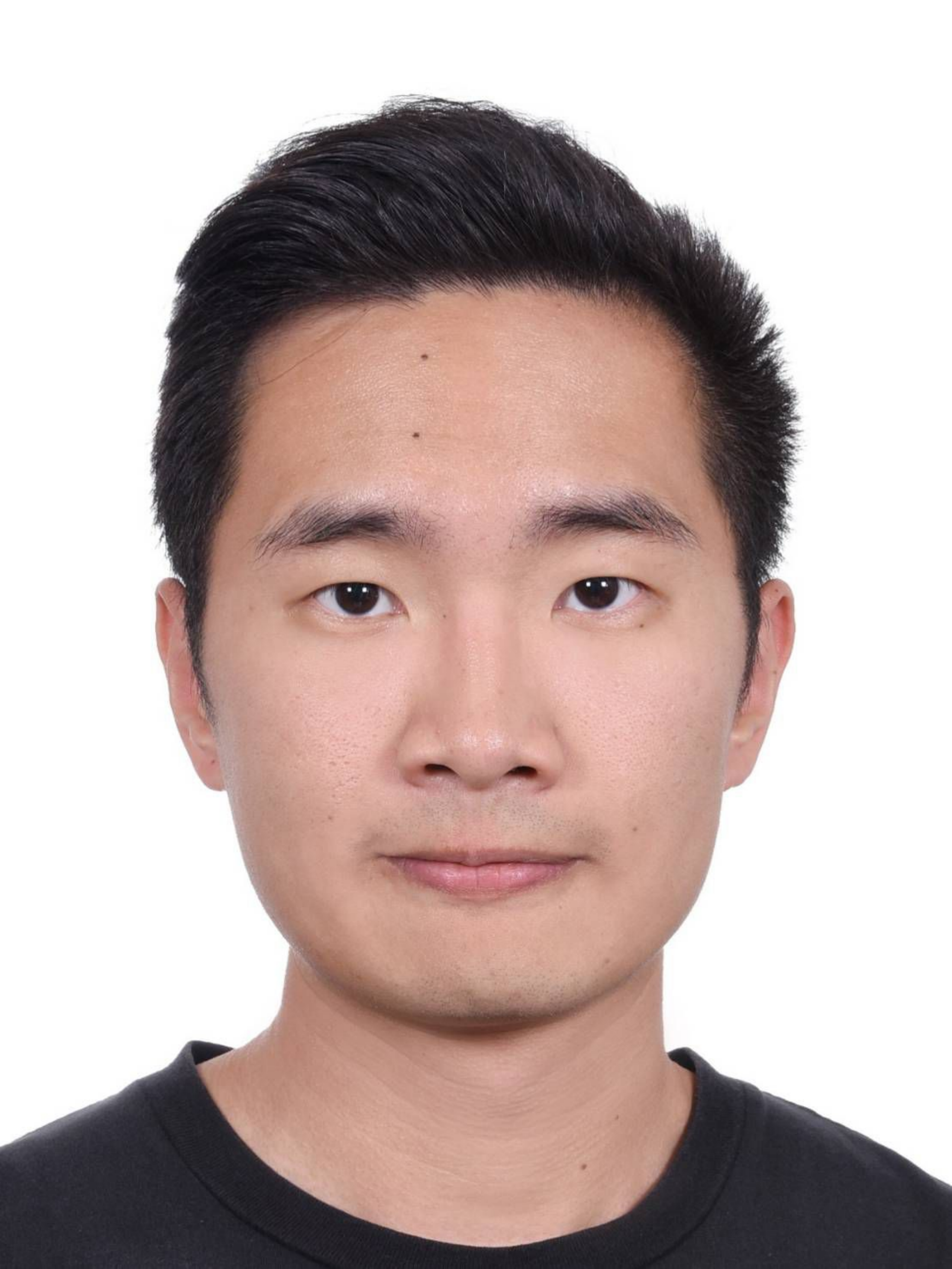}}]{Yu-Huai Peng} received the M.S. degree in Dept. of Electrical  Engineering, National Tsing Hua University, Hsinchu, Taiwan. He is a Research  Assistant in Institute of Information Science, Academia Sinica, Taipei, Taiwan. His research interests include speech signal processing, particularly, voice conversion, and speaker diarization.
\end{IEEEbiography}

\begin{IEEEbiography}[{\includegraphics[width=1in,height=1.25in,clip,keepaspectratio]{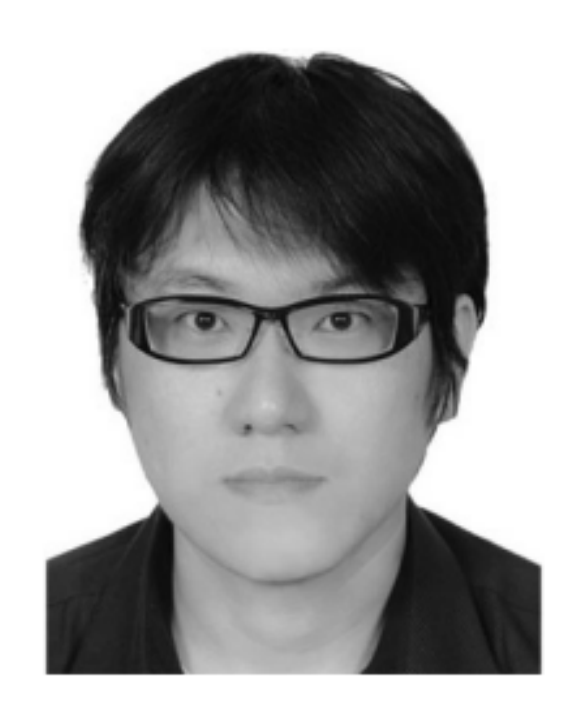}}]{Yu Tsao} (M'09) received the B.S. and M.S. degrees in electrical engineering from National Taiwan University, Taipei, Taiwan, in 1999 and 2001, respectively, and the Ph.D. degree in electrical and computer engineering from the Georgia Institute of Technology, Atlanta, GA, USA, in 2008. From 2009 to 2011, he was a Researcher with the National Institute of Information and Communications Technology, Tokyo, Japan, where he engaged in research and product development in automatic speech recognition for multilingual speech-to-speech translation. He is currently an Associate Research Fellow with the Research Center for Information Technology Innovation, Academia Sinica, Taipei, Taiwan. His research interests include speech and speaker recognition, acoustic and language modeling, audio coding, and bio-signal processing. He is currently an Associate Editor of the IEEE/ACM Transactions on Audio, Speech, and Language Processing and IEICE transactions on Information and Systems. Dr. Tsao received the Academia Sinica Career Development Award in 2017, National Innovation Award in 2018 and 2019, and Outstanding Elite Award, Chung Hwa Rotary Educational Foundation 2019-2020.
\end{IEEEbiography}

\begin{IEEEbiography}[{\includegraphics[width=1in,height=1.25in,clip,keepaspectratio]{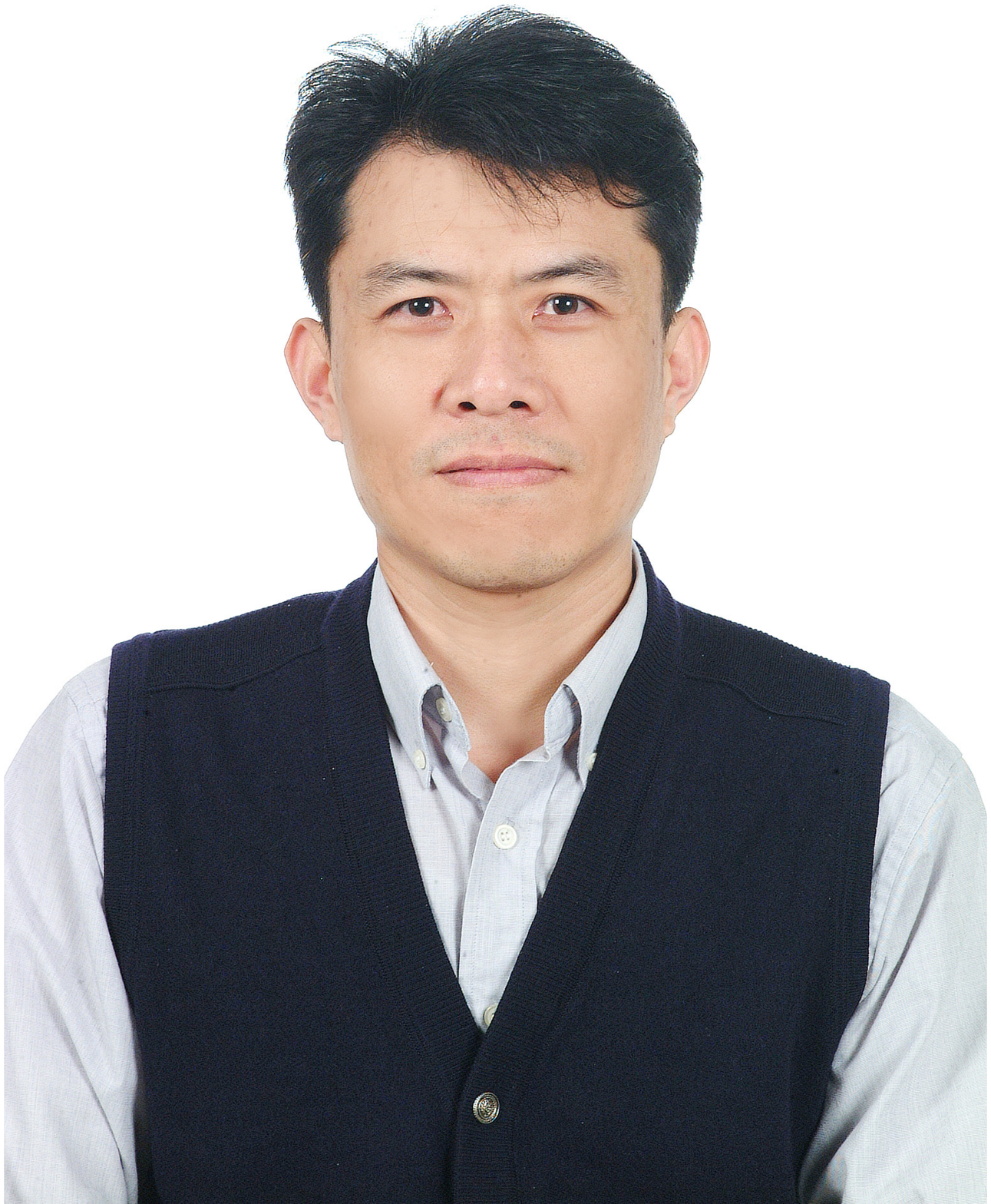}}]{Hsin-Min Wang} (S'92-M'95-SM'04) received the B.S. and Ph.D. degrees in electrical engineering from National Taiwan University, Taipei, Taiwan, in 1989 and 1995, respectively. In October 1995, he joined the Institute of Information Science, Academia Sinica, Taipei, Taiwan, where he is currently a Research Fellow. He also holds a joint appointment as a Professor in the Department of Computer Science and Information Engineering at National Cheng Kung University. He currently serves an Editorial Board Member of IEEE/ACM Transactions on Audio, Speech and Language Processing and APSIPA Transactions on Signal and Information Processing. His major research interests include spoken language processing, natural language processing, multimedia information retrieval, machine learning and pattern recognition. He was a General Co-Chair of ISCSLP2016 and ISCSLP2018 and a Technical Co-Chair of ISCSLP2010, O-COCOSDA2011, APSIPAASC2013, ISMIR2014, and ASRU2019. He received the Chinese Institute of Engineers Technical Paper Award in 1995 and the ACM Multimedia Grand Challenge First Prize in 2012. He was an APSIPA distinguished lecturer for 2014-2015. He is a member of the International Speech Communication Association and ACM.
\end{IEEEbiography}


\vfill


\end{document}